\documentclass[10pt]{article}

\usepackage{amsthm,amsmath}
\usepackage[utf8]{inputenc} 
\usepackage{verbatim}
\usepackage{url}
\usepackage{subcaption}
\usepackage{graphicx}
\usepackage{pdflscape}
\usepackage[margin=0.5in]{geometry}

\begin{document}


\begin{center} Multi-year optimization of malaria intervention: a mathematical model \\
Harry J. Dudley, Abhishek Goenka, Cesar J. Orellana, Susan E. Martonosi \\
For correspondence: Susan E. Martonosi, Ph.D. \\
Harvey Mudd College \\
martonosi@g.hmc.edu 
\end{center}
\begin{abstract} 
\textbf{Background} 

Malaria is a mosquito-borne, lethal disease that affects millions and kills hundreds of thousands of people each year, mostly children.  There is an increasing need for models of malaria control. In this paper, we develop a model for allocating malaria interventions across geographic regions and time, subject to budget constraints, with the aim of minimizing the number of person-days of malaria infection.

\textbf{Methods} 

The model considers a range of several conditions: climatic characteristics, treatment efficacy, distribution costs, and treatment coverage. We couple an expanded susceptible-infected-recovered (SIR) compartment model for the disease dynamics with an integer linear programming (ILP) model for selecting the disease interventions. Our model produces an intervention plan for all regions, identifying which combination of interventions, with which level of coverage, to use in each region and year in a five-year planning horizon.

\textbf{Results}

Simulations using the model yield high-level, qualitative insights on optimal intervention policies: The optimal intervention policy is different when considering a five-year time horizon than when considering only a single year, due to the effects that interventions have on the disease transmission dynamics.  The vaccine intervention is rarely selected, except if its assumed cost is significantly lower than that predicted in the literature.  Increasing the available budget causes the number of person-days of malaria infection to decrease linearly up to a point, after which the benefit of increased budget starts to taper.  The optimal policy is highly dependent on assumptions about mosquito density, selecting different interventions for wet climates with high density than for dry climates with low density, and the interventions are found to be less effective at controlling malaria in the wet climates when attainable intervention coverage is 60\% or lower.  However, when intervention coverage of 80\% is attainable, then malaria prevalence drops quickly in all geographic regions, even when factoring in the greater expense of the higher coverage against a constant budget. 

\textbf{Conclusions}

Our model provides a qualitative decision-making tool to weigh alternatives and guide malaria eradication efforts.  We find that a one-size-fits-all campaign is not cost-effective, and that it is better to consider geographic variations and changes in malaria transmission over time when determining intervention strategies.
\end{abstract}




\section*{Background}

Malaria remains a lethal disease affecting an estimated 200 million people and killing 627,000 in 2012 \cite{WHO2014}.  There are a variety of interventions for treating or preventing malaria infection, but the use of these interventions is hindered by scarcity of resources.    Mathematical models provide a useful tool for evaluating intervention strategies and studying the relative effectiveness of interventions.  These evaluations will become increasingly useful as success with malaria elimination is predicted to change transmission dynamics.  In fact, the WHO Global Malaria Programme cites the specific need for operations research models to determine the best intervention strategies in areas where transmission dynamics are changing as malaria is being eliminated \cite{planningWHO-2013}.  

In this paper, we develop an integer linear program (ILP) and a coupled susceptible-infected-recovered (SIR) compartment model to create a decision-making tool for planning future interventions.  The model suggests the best strategy for minimizing person-days of malaria infection over a five-year period given an initial population, cost of each intervention, and a budget constraint.  We pay special attention to the possibility of a malaria vaccine in combination with other interventions.  We also perform simulations in which we vary our budget, the efficacy of the interventions, and their cost to determine the sensitivity of the optimal policy to these parameters.

\subsection*{Interventions} 

There are many existing methods to prevent or treat malaria infection.  Our model will consider the following five interventions and their combinations.

\textit{Long-lasting insecticidal nets} (LLINs) cover sleeping individuals during the night when mosquito biting can be highest.  When intact, the nets block mosquitoes from reaching humans.  The insecticides work by deterring mosquitoes from feeding and by killing female mosquitoes that come in contact with the net.  LLINs can remain effective for multiple years \cite{CDC2014}.  In fact, the WHO Pesticide Evaluation Scheme 2005 guidelines state that LLINs should survive at least 3 years of recommended washing and use \cite{whopes2005}.

\textit{Indoor residual spraying} (IRS) is another insecticidal prevention method.  IRS is believed to deter mosquitoes from entering sprayed areas and to kill female \textit{Anopheles} mosquitoes that rest on sprayed surfaces after feeding.  (Resting after feeding is a hallmark of some mosquito species while others prefer to rest outdoors \cite{pluess2010}.)  Historically, IRS with an insecticide called dichlorodiphenyltrichloroethane was effective in reducing malaria in Europe, Asia, and Latin America.  However, as insecticide use increases, insecticide resistance has been observed in some mosquito populations in Africa, and new insecticides must be used \cite{WHO2014}.

\textit{Intermittent preventive therapy} (IPT) is the regular administration of a drug like sulfadoxine-pyrimethamnine to decrease morbidity due to malaria in infants, children, and pregnant women.  IPT decreases the chance of developing symptoms after being bitten by an infected mosquito \cite{aponte2009}.  There is evidence that children withstand acute infection better than adults.  However, in endemic areas, adults develop acquired immunity from repeated exposures, and children remain more susceptible to high levels of parasitemia (parasite density in the blood) \cite{doolan2009}.  Most of the 627,000 people killed by malaria in 2012 were children in Africa, so giving IPT to infants, children, and pregnant women  treats the most vulnerable population while limiting the risk of spreading drug resistance \cite{WHO2014}.

\textit{Artemisinin combination therapy} (ACT) can be used to treat a patient after they contract malaria.  This is the best treatment for uncomplicated \textit{P. falciparum} malaria when confirmed by rapid diagnostic tests (RDT) \cite{WHO2014,MC2014}.  ACT kills the parasites that cause symptoms and may destroy or disable the gametocytes that are responsible for infecting mosquitoes \cite{duffy2006}.  Both these factors mean that ACT increases the recovery rate.

Many \textit{malaria vaccines} are in development, and one has gone through Phase III clinical trials.  The complex life-cycle of the malaria parasite makes it possible to intervene at many stages.  Vaccines that target different forms of the parasite will operate by different mechanisms, but in general, a vaccine would decrease the chance of developing symptoms and increase the recovery rate if infected.  The leading malaria vaccine candidate is the RTS,S malaria vaccine.  It is an antigen composed of the RTS and S proteins. The RTS,S vaccine is a pre-erythrocytic vaccine that presents circumsporozoite protein (CSP) from malaria sporozoites to the immune system.  CSP is a parasitic surface protein that is an important part of the invasion of hepatocytes by sporozoites \cite{crompton+pierce+miller2010}.  Such a vaccine will decrease the probability that a susceptible person becomes infected after a bite from an infectious mosquito.  Moreover, it is believed the vaccine could increase a person's recovery rate by increasing their exposure to asexual blood-stage parasites, thereby boosting their immunity \cite{crompton+pierce+miller2010}.  (By contrast, a \textit{transmission-blocking} vaccine that acts in mosquitoes would decrease the probability of transmission from an infectious mosquito but would not change the human recovery rate.)

\subsection*{Literature Review}

Our work extends a single-stage optimization model of Dimitrov \textit{et al.}  Their model divides the country of Nigeria into approximately 270,000 cells and chooses one action (either a single intervention or no intervention) for each cell over a year, subject to budget constraints, to minimize societal costs caused by malaria infection. The model also identifies optimal locations for supply distribution centers.  They treat the societal benefit of each intervention as an exogenous parameter that depends on geographic characteristics.  This allows their model to consider geographic variability in malaria dynamics \cite{dimitrov}. 

However, because malaria dynamics depend on the fraction of the population that is infectious, a quantity that the interventions are themselves trying to reduce, the framework of Dimitrov \textit{et al.} does not permit the examination of multiyear efforts against malaria in which the optimal policy might vary over time as the malaria dynamics shift. This paper extends the optimization model above to select interventions (or combinations thereof) over multiple years by explicitly incorporating malaria disease dynamics over time in response to those interventions.  This is a novel approach that combines two areas of mathematics that do not regularly interact:  integer linear programming from the area of operations research and differential equations modeling from the area of mathematical epidemiology.

There is a long history of mathematical models of malaria transmission, going back to the work of Sir Ronald Ross in the early 1900's \cite{ross1915, mandal+sarkar+sinha2011}. In recent years, malaria has drawn significant attention from the academic community. Epidemiologists have traditionally modeled the spread of malaria in a population using variations on the susceptible-infected-recovered (SIR) model to capture different aspects of the disease.  Mandal \textit{et al.} survey the models found in the literature and offer a hierarchy based on model complexity \cite{mandal+sarkar+sinha2011}.

In order for our model to make informed choices about which interventions to distribute, we must understand how the dynamics of disease transmission change after treatment interventions.  Lindblade \textit{et al.} and Killeen \textit{et al.} study the protective effect of insecticide-treated nets or long-lasting insecticidal nets \cite{lindblade+all2005, killeen+all2006}.  Bousema \textit{et al.} investigate how artemisinin combination therapy reduces the circulation time of gametocytes, thereby reducing infectiousness \cite{bousema+all2010}.  Garner and Graves examine the community benefits of artemisinin combination therapy \cite{garner+graves2005}.  Chandramohan \textit{et al.}, Grobusch \textit{et al.}, and Aponte \textit{et al.} quantify the protective effects of intermittent preventive therapy for infants \cite{ aponte2009, chandrahoman2005, grobusch2007}.  Pluess \textit{et al.} review the effects of indoor residual spray of insecticide \cite{pluess2010}.   These results are used to inform our choice of disease transmission parameters, as described later under \textit{Effects of Interventions on SIR Parameters}.

Our model includes in its portfolio of interventions a vaccine that is currently in development. Prosper \textit{et al.}  model the interaction between vaccine- and naturally-acquired immunity using a five-compartment model.  Their model augments the $S$, $I$, and $R$ classes with a partially-immune (due either to vaccination or natural immunity) susceptible class and a moderately-infectious class for infected, partially-immune individuals.  They find that disease burden can be decreased only if a highly effective vaccine is coupled with a policy of actively treating asymptomatic infections in partially immune individuals \cite{prosper+more2014}. Bojang \textit{et al.} report there is minimal potential effect for a  malaria vaccine given to adult men, and Asante \textit{et al.} study the positive potential protective benefits of administering the vaccine to children \cite{bojang2001, asante2011}.

There are extensions to the SIR framework that we do not consider here.  Koella and Antia model the reduced efficacy of interventions due to the spread of drug-resistant strains of malaria \cite{koella+antia2003}.  Our model does not incorporate drug-resistance, so any policy recommended by our model should be evaluated in this context.  Other researchers, for example Dawes \textit{et al.} \cite{dawes+more2009} and Koudou \textit{et al.} \cite{koudou+more2014}, focus on the mosquitoes' malaria dynamics by analyzing the effects of interventions on mosquito morbidity and mortality rates and the usefulness of the resulting manipulation of said rates. We do not model changing mosquito population explicitly; instead we represent the effects of interventions on the mosquito population as changes in the parameter values used in the human SIR model.

While the above references provide detailed models of malaria's complex dynamics, we have chosen to develop a simple (SIR) model that accommodates the effects of several types of interventions, while maintaining the computational tractability required by the optimization model.  In the next section, we describe our model and simulation approach in greater detail.

\section*{Methods}

We consider the problem of allocating malaria treatments to many regions when limited by scarce resources.  We assume a fixed annual budget shared across several geographic regions having different initial incidences and transmission rates of malaria and different unit costs for distributing treatment.  We consider a portfolio of interventions that can be selected, including some  in combination, each having its own effects on malaria transmission.  Each intervention is selected at a particular \textit{coverage}, which is the percentage of the population that receives the intervention and uses it correctly.  We assume social and economic losses are proportional to the time spent infectious, so  person-days of malaria infection is our chosen measure of the malaria burden.    We seek the optimal sequence of interventions and corresponding coverage percentages for each region and each year that minimizes the total infected person-days over a fixed time horizon.

An integer linear programming optimization model (ILP) suggests the best set of interventions in each year to minimize person-days of malaria infection over all time steps. The ILP takes as input the number of person-days of malaria infection that occur when a given intervention is used on a population with a given initial prevalence of malaria. The person-days of malaria infection is estimated by a susceptible-infected-recovered (SIR) differential equations model of malaria transmission dynamics.

\subsection*{Integer Linear Programming (ILP) Model} 
\label{sec:ILP}

Our ILP relies on several sets, parameters, and decision variables, which we will first define.

\subsubsection*{Sets}
\noindent {Geographic regions:} Because the cost of distributing an intervention to a particular district depends on its infrastructure and ease of access to treatment, and the malaria transmission dynamics depend on its climate, we group districts into geographic regions, denoted by index $g$.  The optimization model determines the number of districts in each geographic region to receive a particular sequence of interventions.  \\

\noindent {Population states: } A population state, $p$, is a triplet, $(S,I,R)$, that indicates the percentage of a district's population susceptible to ($S$), infected by ($I$), or recovered from and temporarily immune to ($R$) malaria. Each district begins a year in a particular population state and ends in a new population state that depends on how the chosen intervention affects the malaria disease dynamics. (The model for determining the disease progression is described in the section \textit{Differential Equations (DE) Model}.)  \\

\noindent {Actions: } The set of actions is the set of possible choices of intervention (including certain combinations of interventions, or the possibility of applying no intervention.) We refer to our choice of intervention at a determined coverage level in a district as an action, denoted by index $i$.    \\

\subsubsection*{Parameters}
$A^{in}_{igpq}$ is an indicator variable whose value is $1$ if action $i$ applied to a district of geographic region $g$, initially in population state $p$ causes a transition to population state $q$, and $0$ otherwise.  
\\
$A^{out}_{igp}$ is an indicator variable whose value is $1$ if action $i$ applied to a district of geographic region $g$, initially in population state $p$ causes a transition to a different population state, and $0$ otherwise. 
\\
$B_{t}$ is the annual budget for year $t$; the combined cost of our actions across all districts in year $t$ must not exceed this value.  \\
$C_{ig}$ is the cost of action $ i $ in any district in geographic region $ g $. \\
$I_{pg}$ is the number of districts in geographic region $g$ that are initially in population state $ p $ at the first time step. \\
$L_{ipg}$ is the number of person-days of malaria infection incurred in a district in geographic region $g$, initially in population state $p$, under action $i$.\\
$N$ is a number larger than the total number of population states. \\
$T$ is the time horizon, in years, considered by the model.\\

\subsubsection*{Decision variables} 
$P_{pgt}$ is the number of districts in geographic region $g$ that are initially in population state $p$ at the start of year $t$.\\
$a^{OUT}_{ipgt}$ is the number of districts in geographic region $g$ that are initially in population state $p$ at the start of year $t$ and are assigned action $i$. \\
$a^{IN}_{ipqgt}$ is the number of districts in geographic region $g$ that are initially in population state $p$ at the start of year $t$, are assigned action $i$, and end in population state $q$. \\

\subsubsection*{Model}
Using these sets, parameters and decision variables, we can define the following integer linear program.
\begin{align}
min &\sum_{t,i,p,g} L_{ipg} a^{OUT}_{ipgt} \label{eq:obj}\\
s.t. &\sum_{i,g} C_{ig} \sum_{p} a^{OUT}_{ipgt} \leq B_{t}& &\forall t \label{eq:c1}\\ 
& P_{p,g,t=1} = I_{pg}& &\forall p, g \label{eq:c2} \\
&\sum_{p}P_{pgt} = \sum_{p}P_{p,g,t+1}& &\forall g, t \label{eq:c3} \\
&\sum_{i}a^{OUT}_{ipgt} = P_{pgt}& &\forall p, g, t \label{eq:c4} \\
& P_{pgt} + \sum_{i,q} a^{IN}_{ipqgt} - \sum_{i} a^{OUT}_{ipgt} = P_{p,g,t+1}& &\forall p, g, t \label{eq:c5} \\
&\sum_{p} a^{IN}_{ipqgt} = a^{OUT}_{iqgt}& &\forall i, q, g, t \label{eq:c8} \\ 
& a^{IN}_{ipqgt} \leq N \cdot A^{in}_{igpq}& &\forall i, p, q, g, t \label{eq:c6} \\  
& a^{OUT}_{ipgt} \leq N \cdot A^{out}_{igp}& &\forall i, p, g, t \label{eq:c7} \\
& P_{pgt}, a^{IN}_{ipqgt}, a^{OUT}_{ipgt} \geq 0, integer& &\forall i, p, q, g, t \nonumber 
\end{align}

The objective function in expression (\ref{eq:obj}) minimizes the cumulative person-days that each district spends in the infected state over the time horizon, as a function of our choice of actions.  Constraint \eqref{eq:c1} requires the chosen set of  interventions to be within budget in each year.
Constraint \eqref{eq:c2} initializes the population variable at the start of the time horizon.
Constraints \eqref{eq:c3}---\eqref{eq:c8} are bookkeeping constraints that keep track of the number of districts in each geographic region and population state as a function of the actions chosen.  
Constraints \eqref{eq:c6} and \eqref{eq:c7} assure that districts transition out of population state $p$ to population state $q$ only when an appropriate action has been taken.
The last constraint requires  all decision variables to be nonnegative integers.

\subsection*{Differential Equations (DE) Model}
\label{sec:genericDE}
Several of the parameters used by the ILP model, specifically $A^{in}_{igpq}$, $A^{out}_{igp}$ and $L_{ipg}$ depend on the dynamics of malaria progression.  The susceptible-infected-recovered (SIR) model is a standard system of nonlinear ordinary differential equations for analyzing the transmission of malaria \cite{mandal+sarkar+sinha2011,koella+antia2003}.  We modify the standard model and use a coupled six-class compartment model with separate SIR compartments for treated and untreated individuals.  This coupling of treated and untreated SIR classes allows us to model population-wide benefits caused by decreased infectiousness of a treated subpopulation.     For an initial population state and action, we solve this system of equations to determine the population state after one year.  This gives us the indicator parameters $A^{in}_{igpq}$ and $A^{out}_{igp}$. We also use the solution to this system of differential equations to estimate the burden of malaria, measured in infected person-days, during that year.  For each district in geographic region $g$, beginning the year in a particular population state $p$, having been assigned action $i$, we numerically integrate the infected class curve that results under those  conditions, multiplied by the district's population.  This estimates the number of people who are infected over the year times the number of days for which they remain infected.  This number is then input into the linear programming model as the value of $L_{ipg}$. This is pre-solved for all possible population states and actions, and the results are stored as input data for the ILP.

We now define the parameters, state variables and system of differential equations.

\subsubsection*{Parameters} 
    $a_u$ ($a_t$) is the number of bites per mosquito per untreated (respectively, treated) human per day. \\
    $b_u$ ($b_t$) is the transmission efficacy from infected mosquito to susceptible, untreated (resp., treated) human. \\
    $c$ is the transmission efficacy from infected human to susceptible mosquito. \\
    $\delta$ is the daily birth rate and death rate.  We assume constant population. \\ 
    $\gamma_u$ ($\gamma_t$) is the recovery rate for untreated (resp., treated) people.  Its reciprocal is the average time that a person is infected with malaria.  \\ 
    $h_u$ ($h_t$) is the \textit{force of infection}, that is, the rate at which untreated (resp., treated) susceptible humans become infected with malaria. \\
    $m_u$ ($m_t$) is the number of mosquitoes per untreated (resp., treated) human. \\
    $\mu$ is the mosquito mortality rate. \\ 
    $\omega$ is the duration of immunity without reinfection. \\
    $q$ is the treatment coverage, the percentage of the population that receives a treatment and uses it correctly.  We assume the same percentage of newborns are born into the susceptible, treated class.  The remaining fraction, $1-q$, are born into the susceptible, untreated class. \\
    $\rho_u$ ($\rho_t$) is rate of immunity loss for recovered untreated (resp., treated) humans. \\
    $\tau$ is the incubation period of malaria in the mosquito. \\

\subsubsection*{State Variables}
$S_u$ ($S_t$) is the proportion of the population that is susceptible and untreated (resp., treated). \\
$I_u$ ($I_t$) is the proportion of the  population that is symptomatic,  infectious, and untreated (resp., treated). \\
$R_u$ ($R_t$) is the proportion of the  population that is recovered with acquired immunity and untreated (resp., treated). \\

\subsubsection*{Model} 
The proportions of the population belonging to each of the six classes can be determined by solving the following system of differential equations:
\begin{align} 
&\frac{dS_u}{dt} = \delta(1-q) - (\delta + h_u)S_u + \rho_u R_u \label{eq:general_dSudt} \\
&\frac{dI_u}{dt} = h_u S_u - (\delta + \gamma_u )I_u \label{eq:general_dIudt} \\
&\frac{dR_u}{dt} = \gamma_uI_u - (\delta + \rho_u)R_u \label{eq:general_dRudt} \\
&\frac{dS_t}{dt} = \delta q - (\delta + h_t)S_t + \rho_t R_t \label{eq:general_dStdt} \\
&\frac{dI_t}{dt} = h_t S_t - (\delta + \gamma_t)I_t  \label{eq:general_dItdt} \\
&\frac{dR_t}{dt} = \gamma_t I_t - (\delta + \rho_t)R_t. \label{eq:general_dRtdt}
\end{align}

Although on the surface, the equations for the untreated population and the equations for the treated population do not appear to be coupled, the coupling occurs with the parameters $h_u$ and $h_t$, which are the force of infection parameters.  They have been derived by Smith and McKenzie \cite{smith+mckenzie2004} to be:
\begin{equation} \label{eq:h_u}
  h_u = \frac{m_u a_u^2 b_u c e^{-\mu \tau} (I_u + I_t)}{\mu + a_u c (I_u + I_t)} 
\end{equation}
\begin{equation} \label{eq:h_t}
  h_t = \frac{m_t a_t^2 b_t c e^{-\mu \tau} (I_u + I_t)}{\mu + a_t c (I_u + I_t)} .
\end{equation}  We see these rates are functions of the total proportion of infectious people, $I_u + I_t$, which couples the system of differential equations.  The more infectious people there are in either the untreated or treated group, the faster the rate at which susceptible people in either group can become infected.

The rates of immunity loss, $\rho_u$ and $\rho_t$, are functions of $h_u$ and $h_t$, respectively and further couple the system.  The procedure for deriving the rate of immunity loss has been shown by Aron and May \cite{aron+may1982}.  These equations assume that being exposed to malaria while recovering resets the duration of immunity. 
\begin{equation}  \label{eq:rho_u} 
  \rho_u = \frac{h_u+\delta}{e^{\omega(h_u+\delta)}-1} 
\end{equation}
\begin{equation}  \label{eq:rho_t} 
  \rho_t = \frac{h_t+\delta}{e^{\omega(h_t+\delta)}-1} 
\end{equation}

This is a general model that does not consider the effect an intervention can have on the transmission of the disease.  In the specific case of ACT, a medication that clears infection rapidly, the length of time a malaria patient is carrying infectious gametocytes in her blood is significantly reduced \cite{bousema+all2010, sutherland+all2005}.  Because of this, we make the assumption that ACT clears parasites before the body has time to develop acquired immunity; therefore, we assume infected people treated with ACT skip the recovered class and transition directly back to the susceptible class. To reflect this, we introduce the indicator variable $\psi_{act}$, which equals $1$ when ACT is chosen (either alone or in combination with another intervention), and $0$ otherwise.  The state equations for the untreated class are unchanged, and the equations for the treated classes become:

\begin{align} 
&\frac{dS_t}{dt} = \delta q - (\delta + h_t)S_t + \rho_t R_t + \psi_{act} \gamma_t I_t \label{eq:dStdt} \\
&\frac{dI_t}{dt} = h_t S_t - (\delta + \gamma_t)I_t  \label{eq:dItdt} \\
&\frac{dR_t}{dt} = (1-\psi_{act})\gamma_t I_t - (\delta + \rho_t)R_t \label{eq:dRtdt}
\end{align}
We see that when ACT is used, infectious individuals bypass the recovered class and transition directly to the susceptible class. 

Because we are selecting a new portfolio of interventions to distribute each year, we assume the effects of treatment last for one year, exactly.  Some of the treatments are known to last longer; for instance, the insecticide coating on mosquito nets is believed to be effective for three years, and vaccines in development currently have an efficacy of three years.  However, assuming a duration of only one year is conservative: under this assumption, our model will underestimate the efficacy of our interventions, and the results we would expect to see in the field should be better.  Under this assumption, at the end of each year, the six-state population $(S_u, I_u, R_u, S_t, I_t, R_t)$ can be collapsed into a more compact three-state representation:  $(S_u+S_t, I_u+I_t, R_u+R_t)$. 

\subsubsection*{Coverage} 
The coverage, $q$, refers to the percentage of the population that receives a treatment and uses it correctly.  For example, if at the start of the year, the percentages of the population who are susceptible, infected and recovered are given by $(S, I, R)$, respectively, then the initial values of $S_u, I_u, R_u, S_t, I_t,$ and $R_t$ for the differential equations model will be $(1-q)S, (1-q)I, (1-q)R, qS, qI,$ and $qR$, respectively.  

However, some interventions, such as IPT and vaccine, are assumed to be distributed only to newborns and children under the age of four.  In these cases, the coverage, $q$, applies only to births and to the fraction of the population under the age of four.  If $x$ is the fraction of the population under the age of four, and $(S, I, R)$ is the initial distribution of susceptible, infected and recovered individuals in the population, then the initial values of $S_u, I_u, R_u, S_t, I_t,$ and $R_t$ for the differential equations model will be $(1-qx)S, (1-qx)I, (1-qx)R, qxS, qxI,$ and $qxR$.

\subsection*{Data}

Our model relies on parameters governing intervention costs, malaria transmission, and intervention efficacy.  When available, we estimate parameter values based on malaria research literature.  When using country-specific information, we use data from Kenya or its neighbors as it is more readily available and allows us to be consistent across parameters.  We also conduct sensitivity analysis to understand how the model's results would change under a range of scenarios concerning distribution costs, climate and intervention efficacy.  In this section, we first describe the costs of the interventions, then we describe the baseline parameter values used in the $SIR$ model, and we outline the changes in these parameter values under interventions and sensitivity analysis scenarios.

\subsubsection*{Base Costs of Interventions}
Our model parameter $C_{ig}$ is the per person, per year cost of action $ i $ in any district in geographic region $ g $.  The cost of an action depends on the purchase price as well as transportation and distribution costs, which we assume are regional.  For the base cost, we used data provided by White \textit{et al.}, who survey cost and cost-effectiveness data for LLIN, IRS, IPT, and ACT from all available sources and adjust it to 2009 USD \cite{white2011}.  We focused specifically on data from Kenya, except where noted that no Kenya-specific data was available; in these cases, we used cost estimates from nearby Ethiopia, Tanzania and Zimbabwe.

First, we list the base cost for each intervention, and in the subsequent section, we will describe how we modify those costs to reflect transportation and distribution costs in different geographic regions.  These are summarized in Table \ref{tab:costs}. \\

\noindent {LLIN:} The average cost of a single insecticide-treated mosquito net is 7.21 USD \cite{white2011}, and the WHO Pesticide Evaluation Scheme 2005 guidelines estimate a three year life span with recommended use  \cite{whopes2005}.  Because our model assumes all actions expire at the end of one year, we use an annual cost per net of 2.40 USD, which is one-third the base cost of the net.  Moreover, bed-sharing is a common practice that further reduces the per-person cost of each distributed net.  The World Health Organization recommends the assumption that an LLIN will protect 1.8 people on average \cite{WHO2014nets}, making the annual per-person cost 1.33 USD.    \\  

\noindent {IRS:} Our IRS cost estimate assumes two rounds of household spraying with lambdacyhalothrin per person per year, at an annual cost of 2.22 USD \cite{white2011}.  \\

\noindent {IPT:} White \textit{et al.} summarize cost estimates for distributing IPT to newborns, children and pregnant women.  The mean  cost of distributing six bi-monthly doses of sulfadoxine-pyrimethamine to infants in Tanzania is reported to be 0.78 USD, and three doses per year to children in Kenya is 1.25 USD \cite{white2011}.  As roughly 25\% of children under the age of 4 are infants, we can estimate a weighted average annual cost for IPT of 1.13 USD.  \\

\noindent {ACT:} White \textit{et al.} report malaria diagnosis and treatment costs for a variety of diagnostic methods and treatment types in several countries.  For consistency, we used costs associated specifically with rapid diagnostic tests (RDT) used in conjunction with ACT treatment in the countries of Tanzania and Zambia.  These ranged from 3.63 USD to 6.72 USD, with an average of 4.82 USD per person treated \cite{white2011}.  Unlike interventions such as long-lasting insecticidal nets, which are assumed to be distributed to the entire treated class, ACT is distributed only to members of the treated class who experience a malaria infection.  Therefore, our SIR model must estimate the number of new malaria infections per year to determine the annual cost of ACT.  According to equation (\ref{eq:dItdt}), new infections occur with rate $h_tS_t = \frac{dI_t}{dt} +(\delta+\gamma_t)I_t$.  Note that $\frac{dI_t}{dt}$ can be approximated by $\frac{I_t(d+\epsilon)-I_t(d)}{\epsilon}$ for small $\epsilon$.  If we discretize the year over which the treatment is available into 365 days and let $\epsilon=1$ day, we find that the number of new infections appearing on day $d$ should be roughly $I_t(d) - (1-(\delta+\gamma_t))I_t(d-1)$ times the total population. Summing this value over all days $d$ should give an approximation of the number of new infections incurred during the year, and hence, the number of people who received ACT. \\

\noindent {Vaccine:} Cost data for the RTS,S vaccine is not yet available since the vaccine is not yet on the market.  Seo \textit{et al.} use an estimate of 7 USD per dose for the vaccine after looking at recent introductory vaccine prices ranging from 1 to 15 USD \cite{seo2014}.  They also propose using 0.37 USD administration cost per vaccination based on the price for other vaccines used in Malawi in the Expanded Program on Immunization (EPI).  Because the RTS,S vaccine is administered in three doses, they estimate the total cost of the vaccine at 22.11 USD per person per year \cite{seo2014}. Adjusting their 2012 costs to 2009 values for consistency, we arrive at a cost of 20.66 USD per treated person per year \cite{BLSCalc}.

\subsubsection*{Baseline SIR Model Parameter Values}
\label{subsubsec:SIRParams}

We now describe the baseline parameter values used in the $SIR$ model and discuss how the interventions and modeling assumptions affect those values in the next section.  This information is summarized in Table \ref{tab:parameters}.

\begin{itemize}
\item $a_u$  is the number of bites per mosquito per untreated human per day, which is estimated to be 0.25 \cite{chitnis+hyman+cushing2008}. 
\item $b_u$  is the transmission probability from infected mosquito to susceptible, untreated  human, which is estimated to be 0.022 \cite{chitnis+hyman+cushing2008}. 
\item $c$ is the transmission probability from infected human to susceptible mosquito, which is estimated to be 0.36 \cite{chitnis+hyman+cushing2008}. 
\item $\delta$ is the daily birth rate and death rate.  In Kenya in 2014, the estimated annual birth rate was $0.02827$ births per person, and the estimated annual death rate was $0.007$ deaths per person \cite{ciafactbook}.  Because we assume a constant population in our model, we use the average of these, or $0.017635$ births (deaths) per person per year, and convert it using compounding to a daily birth (death) rate of $\delta = 4.7895 \times 10^{-5}$.   
\item $\gamma_u$ is the recovery rate for untreated people.  Filipe \textit{et al.} estimate the average infectious period for untreated people to be 180 days, making $\gamma_u = \frac{1}{180}$ \cite{filipe+all2007}.  
\item $m_u$  is the mosquito density (number of mosquitoes per untreated human), which is estimated to be 20 \cite{mandal+sarkar+sinha2011}. 
\item $\mu$ is the mosquito mortality rate, estimated to be 0.095 days$^{-1}$ \cite{filipe+all2007, anderson+may1991}. 
\item  $\omega$ is the duration of immunity without reinfection.  We use the value $\omega = 274$ days, based on an estimate that immunity lasts between six and twelve months \cite{kakkilaya2011}.  
\item $q$ is the treatment coverage, the percentage of the population that receives a treatment and uses it correctly.   We consider three levels of treatment coverage for each intervention, high (60\%), medium (40\%), or low (20\%). 
\item $\tau$ is the incubation period in the mosquito, estimated to be 10 days \cite{anderson+may1991}. 
\item $x$ is the fraction of the population that is age four years or younger, which was approximately 14.6\% in Kenya in 2014 \cite{ciafactbook}. 
\end{itemize}

To determine the force of infection ($h_u$ and $h_t$) and the recovery rate ($\rho_u$ and $\rho_t$), we use the expressions given in equations \eqref{eq:h_u} -- \eqref{eq:rho_t}.

Note that the malaria transmission parameters, $a_u$, $b_u$, $c$, and $m_u$, are very location-specific (see \cite{anderson+may1991}, p. 409). Adapting this model to any location would require re-estimating these parameters.

\subsubsection*{Effects of Interventions on $SIR$ Parameters}
Each intervention, or combination of interventions, is modeled as affecting a subset of the above parameters.  \\

\noindent {LLINs} protect individual users by decreasing the biting rate, $a_t$, and by killing mosquitoes that contact the insecticidal nets, thus decreasing  $m_t$.  We estimate the values of $a_t$ and $m_t$ as follows:
\begin{itemize}
\item Let $\beta$ be the proportion of mosquito exposure that occurs during sleeping hours.   
\item Let $\chi_{LLIN}$ be the probability of mortality for a mosquito exposed to a treated net.  
\item As before, let $m_u$ be the baseline mosquito density absent any treatment.
\end{itemize}
Then Killeen \textit{et al.} \cite{killeen+all2006} derive the value of $a_t$ for people using LLIN as
\begin{equation} \label{eq:LLINa}
a_t = a_u(1-\beta),
\end{equation} and the value of $m_t$ for people using LLIN as
\begin{equation} \label{eq:LLINm}
  m_t = m_u ( 1 - \beta \chi_{LLIN}).
\end{equation}
Although Figure 5 in reference \cite{killeen+all2006} shows a slight increase in overall protection for the treated class as a function of $q$, this increase is modest in the range of $q$ that we consider, and so we assume that $m_t$ is independent of $q$.  Additionally, Killeen \textit{et al.} suggest that as treatment coverage increases in a population, even non-users of LLINs benefit from decreased mosquito density.  However, we were unable to find empirical data to support a robust model of mosquito density in the untreated population as a function of treatment coverage; therefore, we assume the untreated population experiences the baseline mosquito density, $m_u$, for all values of $q$.  

To determine the new biting rate, $a_t$, and the new mosquito density, $m_t$, for the treated classes, we use Equations (\ref{eq:LLINa}) and (\ref{eq:LLINm}), respectively, with $\beta = 0.8$ \cite{killeen+all2006} and $\chi_{LLIN} = 0.8$ \cite{lindblade+all2005}, and with $a_u =  0.25$, and $m_u =20$ as given earlier. \\

\noindent{IRS} decreases the number of mosquitoes per treated human, $m_t$, in a similar manner as LLINs.  Moreover, IRS can also decrease the mosquito density in untreated areas close to treated areas; thus, $m_u$ is also affected by IRS \cite{zhou+all2010}.  We let $\chi_{IRS_t}$ be the reduction in mosquito density in a house treated with IRS, and we let $\chi_{IRS_u}$ be the reduction in mosquito density in an untreated house when the treatment coverage is 100\% in a nearby area.  Then the value of $m_t$ for a house treated with IRS is
\begin{equation} \label{eq:IRSm}
  m_t = m_u ( 1 - \chi_{IRS_t})  
\end{equation}
The value of $m_u$ when IRS is used at coverage $q$ is estimated (based on the empirical results of Zhou \textit{et al.} \cite{zhou+all2010}) to be
\begin{equation} \label{eq:IRSm_u}
  m_{u_{IRS}} = m_u ( 1 - q \chi_{IRS_u}). 
\end{equation}

As \cite{zhou+all2010} report that the mosquito density in treated areas decreases by 95\%, and the mosquito density in untreated areas decreases by 50\% when the coverage in nearby treated areas is 100\%, we use $\chi_{IRS_t} = 0.95$ and $\chi_{IRS_u} = 0.5$,  $m_u = 20$ as given earlier, and $q$ equal to the coverage associated with the selected action. \\

\noindent{IPT} decreases the probability, $b_t$, that a susceptible person becomes infected after a bite from an infectious mosquito. We were unable to find an estimate in the literature for the amount by which the transmission efficacy, $b_t$, decreases when a person is using intermittent preventative treatment.  However, data from several studies reported by Aponte \textit{et al.} indicate that the protective efficacy against malaria in infants of one year of IPT is roughly 30\% \cite{aponte2009}.  As this should roughly correspond to the percentage decrease in new malaria infections observed in the $SIR$ model output, we calibrated the model by solving the system of differential equations for a range of values for $b_t$ and selecting the value of $b_t$ that achieves a 30\% reduction in new malaria infections.  The value $b_t =0.0047$  achieves this percentage reduction.   \\

\noindent{ACT} dramatically  reduces the length of time a malaria patient is carrying infectious gametocytes in her blood, possibly down to a mean infectious period of ten days, so we use $\gamma_t = \frac{1}{10}$ days$^{-1}$ \cite{bousema+all2010, garner+graves2005, sutherland+all2005}.  \\

\noindent{Vaccine}, like IPT, decreases the probability, $b_t$, that a susceptible person becomes infected after a bite from an infectious mosquito.  Additionally, a vaccine could increase the recovery rate, $\gamma_t$, by exposing the immune system to parasite proteins or decreasing the amount of parasites that reach the blood stage initially \cite{crompton+pierce+miller2010}. Olotu \textit{et al.} report clinical trial results suggesting that the four-year reduction in malaria episodes among vaccinated children is 23.5-24.3\% \cite{olotu+all2013}.  As this should roughly correspond to the percentage decrease in new malaria infections observed in the $SIR$ model output, we calibrated the model by solving the system of differential equations for a range of values for both $b_t$ and $\gamma_t$ and selecting the combination that achieves a roughly 24\% reduction in new malaria infections. Choosing $b_t = 0.005$ and $\gamma_t = \frac{1}{5.5}$ days$^{-1}$ achieves this percentage reduction. \\

Possible actions that can be selected by the optimization model are to deploy no intervention, a single intervention, or a combination of two interventions.  Although it is possible to consider combining any pair of interventions, for modeling simplicity, we only consider pairs of interventions whose coverage applies to the same segments of the population.  Therefore, because IPT and vaccines are assumed in our model to be distributed only to newborns and children under the age of four, while LLIN, ACT and IRS are applied to the general population, the combinations we consider are IPT with vaccine,  LLIN with ACT, LLIN with IRS, or ACT with IRS.  

When two interventions are used in combination, we assume the covered segment of the population receives both treatments, and the uncovered segment receives neither.  If the two interventions affect non-overlapping parameter sets, we assume the combination intervention will affect the union of both sets of parameters in the same manner as the individual interventions. However, some pairs of interventions act upon the same parameter.  For example, in the case of LLIN combined with IRS, the mosquito density, $m_t$ is affected by both interventions. Because we think it would be too optimistic to assume that the effects of LLIN and IRS are additive, we make a more conservative assumption: we use the smallest values of $a_t$, $m_u$ and $m_t$ offered by either LLIN or IRS. For the IPT with vaccine combination, $b_t$, the transmission efficacy from infected mosquito to susceptible, treated human, is reduced by both interventions via different mechanisms, and $\gamma_t$ is increased by the vaccine.  We use the smaller of the two $b_t$ values under IPT and vaccine (that given by IPT, of $b_t=0.0047$) and the value of $\gamma_t = \frac{1}{5.5}$ yielded by the vaccine.

\subsubsection*{Sensitivity Analysis Simulations}
We examined the optimal sequence of five-year interventions through simulation of a fictitious nation.  This nation consists of 4,500 districts, each having a population of 10,000 (for a total population of 45 million, comparable to that of Kenya in 2014 \cite{ciafactbook}).  We assume an annual budget of $B_t = 33.75$ million USD, which corresponds to $0.75$ USD per person, per year.  (This is comparable to the budget for the President's Malaria Initiative (PMI) in Kenya, which in 2013 was 34,256,770 USD \cite{pmikenya2013}.)

Each of the 4,500 districts is characterized as belonging to one of nine geographic regions, which in turn are characterized by one of three distribution regions and one of three climate regions. 500 districts belong to each of the nine possible combinations.  Distribution regions categorize districts by how inexpensively (relative to the baseline costs given earlier) interventions can be distributed to the district.  Rural and remote areas are likely to experience higher-than-baseline distribution costs due to having worse road infrastructure and lower access to health centers.  Centrally located urban areas are likely to experience lower-than-baseline distribution costs.  We assume districts categorized as having low distribution costs will have intervention costs that are 20\% lower than the baseline values given in Table \ref{tab:costs}.  Districts categorized as having medium distribution costs will incur the baseline intervention costs, and districts categorized as having high distribution costs will incur intervention costs that are 20\% more than the baseline costs given in Table \ref{tab:costs}.

The three climate regions, dry, moderate and wet, reflect the effect climatic characteristics such as temperature and precipitation can have on mosquito population, and hence on malaria transmission dynamics.  The moderate climate scenario assumes the baseline mosquito density given above of $m_u = 20$ mosquitoes per human  The dry scenario assumes a mosquito density of $m_u = 5$ mosquitoes per human, and the wet scenario assumes a mosquito density of $m_u=35$ mosquitoes per human.  Each region is also characterized by its own initial population distribution amongst the $S, I$ and $R$ classes, which was chosen to be the steady-state population distribution observed when the $SIR$ model is run for a long period of time from a variety of starting population distributions and assuming no intervention.  For the moderate climate scenario, the steady-state distribution used for the initial distribution was 15\% susceptible, 15\% infected, and 70\% recovered.  The dry scenario used an initial steady-state distribution of 60\% susceptible, 15\% infected, and 25\% recovered.  The wet scenario used an initial steady-state population distribution of 10\% susceptible, 15\% infected, and 75\% recovered.  (For computational tractability, we use a population state resolution of five percentiles, and round the $SIR$ population state to the nearest 5\%, while requiring that the percentages over all compartments sum to one.)    

We also consider three efficacy scenarios to test the sensitivity of our model to the inherent uncertainty in our understanding of how effective the interventions are.  The baseline efficacy scenario uses the baseline parameter estimates described earlier and shown in Table \ref{tab:parameters}.  The pessimistic efficacy scenario assumes each parameter value for the treated class is 30\% ``worse'' than its baseline value, where by ``worse'', we mean leading to greater malaria infections.  The optimistic efficacy scenario assumes each parameter value for the treated class is 30\% ``better'' than its baseline value. In the case where making a treatment parameter 30\% worse than its baseline value ends up making it worse than the untreated baseline value, the value is capped at the untreated baseline; in this way, we avoid a situation where the treated class might artificially experience more malaria cases than the untreated class.

\section*{Results and Discussion} 
\label{sec:results}

We now present the results of running our model on the data described above. Because we use a simplified SIR model to estimate the social costs of malaria as a function of interventions distributed, the purpose of this model is not to give exact estimates of reductions in person-days of malaria infection, but relative results that can be used by the optimization model to make choices between the various interventions.    Our model provides qualitative insights about trends in the optimal interventions as certain parameters vary; these qualitative insights can then be used to offer high-level policy recommendations, as we describe. 

\begin{figure}[h!]
   \includegraphics[width=\textwidth]{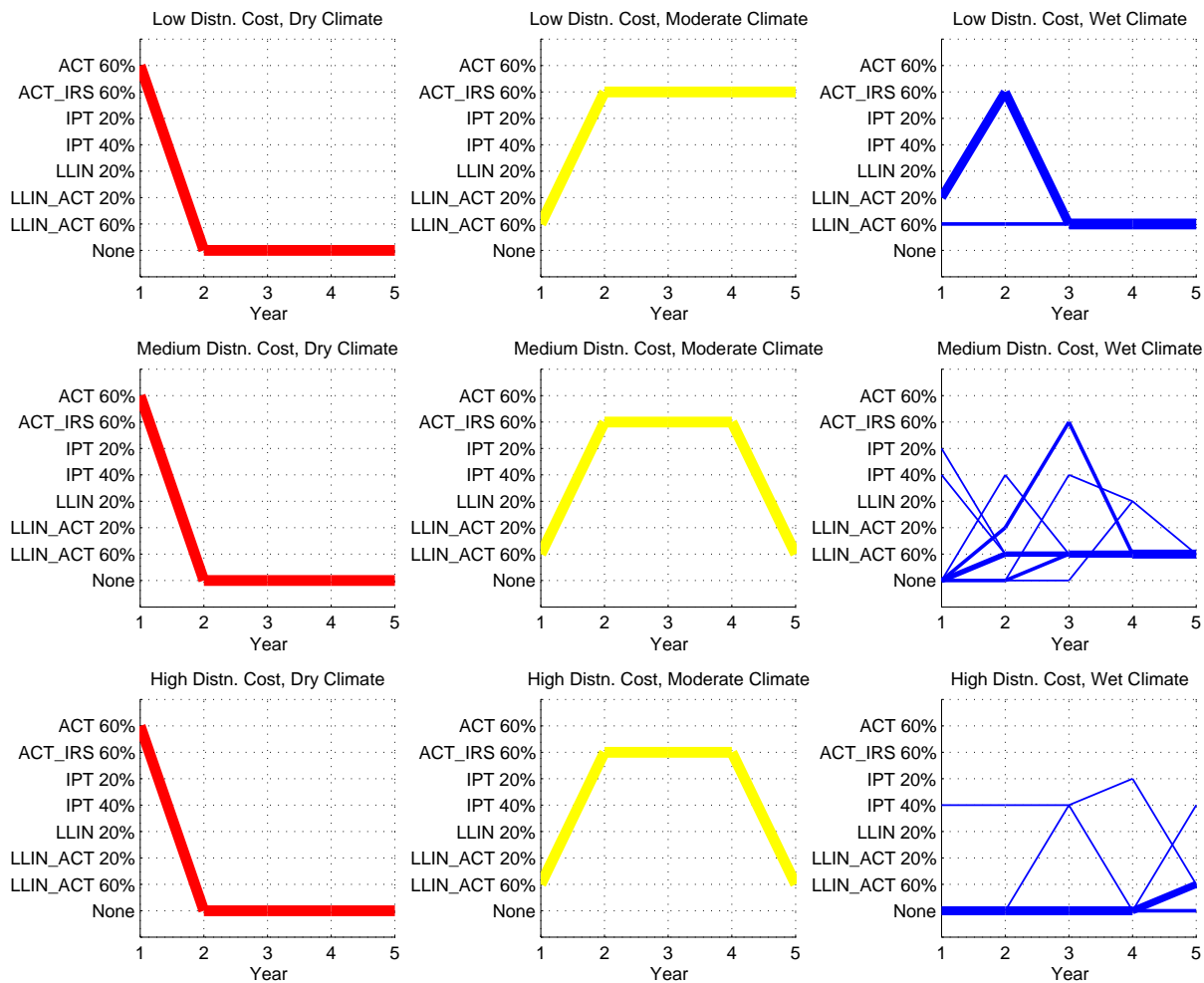}
    \caption{Sequence of interventions over a five-year time horizon in the baseline efficacy scenario.  The thickness of each line is proportional to the number of districts assigned a given sequence.  Intervention sequences assigned to only a single district have been omitted from the figure (5 districts omitted in total).  The total person-days of malaria infection in this scenario is 4.506 billion.}
    \label{fig:baselineoutput}
    \end{figure}

Table \ref{tab:baselineoutput} and Figure \ref{fig:baselineoutput} provide the sequence of interventions allocated in each of the nine geographic regions (low, medium or high distribution costs, crossed with dry, moderate or wet climate) over a five-year horizon in the baseline efficacy case.  The total person-days of malaria infection was found to be 4.506 billion.  In Figure \ref{fig:baselineoutput}, the first row of figures corresponds to low distribution costs, the second row to medium distribution costs and the third row to high distribution costs.  Likewise, each column of figures corresponds to the same climate region (dry in red, moderate in  yellow, wet in blue, from left to right).  The vertical axis on each graph lists in alphabetical order the interventions selected by the model at least once in at least one geographic region and the associated coverage.  The thickness of a path is proportional to the number of districts (out of 500) that were assigned that sequence of interventions.  Table \ref{tab:baselineoutput} gives the number of districts of each type that are assigned a particular sequence of interventions, as well as the resulting $(S,I,R)$ population state after each year.

We see that for dry climate regions, the sequence of interventions is the same regardless of distribution cost, namely, ACT is distributed to 60\% of the population in year 1, and then no subsequent interventions are distributed in years 2-5.  The reason for this is that distributing ACT at a coverage of 60\% in year 1 eradicates (at least subject to rounding at our resolution of 5\%) malaria, driving the infected proportion of the population to zero.  Per equations (\ref{eq:h_u}) and (\ref{eq:h_t}), the force of infection is zero when the infected population is zero, and the disease cannot persist.  

For the moderate climate regions, all districts are assigned ACT combined with LLIN at 60\% coverage in the first year, followed by ACT in combination with either LLIN or IRS at 60\% coverage in subsequent years.  

The sequence of interventions assigned in the wet regions at first glance appears more interesting. First we see that only low distribution cost districts, along with only two medium distribution cost and one high distribution cost districts, receive any intervention during the first year. 126 high distribution cost districts never receive any intervention in any of the five years.  This is likely due to the budget constraint forcing the model to prioritize eliminating malaria in the dry climate regions during year one and leaving the harder-to-access wet regions largely untreated. In those wet districts receiving interventions, the chosen interventions are primarily LLIN with ACT at 60\% coverage.  but as Table \ref{tab:baselineoutput} shows, these interventions do little to reduce the prevalence of malaria in the population. An apparent steady-state consists of 10\% of the population in the infected state even after several years of 60\% coverage of LLIN with ACT.  Thus, we can infer that combating malaria in the wet regions is not possible with the interventions considered at coverage percentages up to 60\%.  As we show later, increasing the maximum coverage to 80\% is necessary for reducing malaria in wet regions.

\subsection*{Effect of Treatment Efficacy}
\begin{figure}[h!]
  \includegraphics[width=\textwidth]{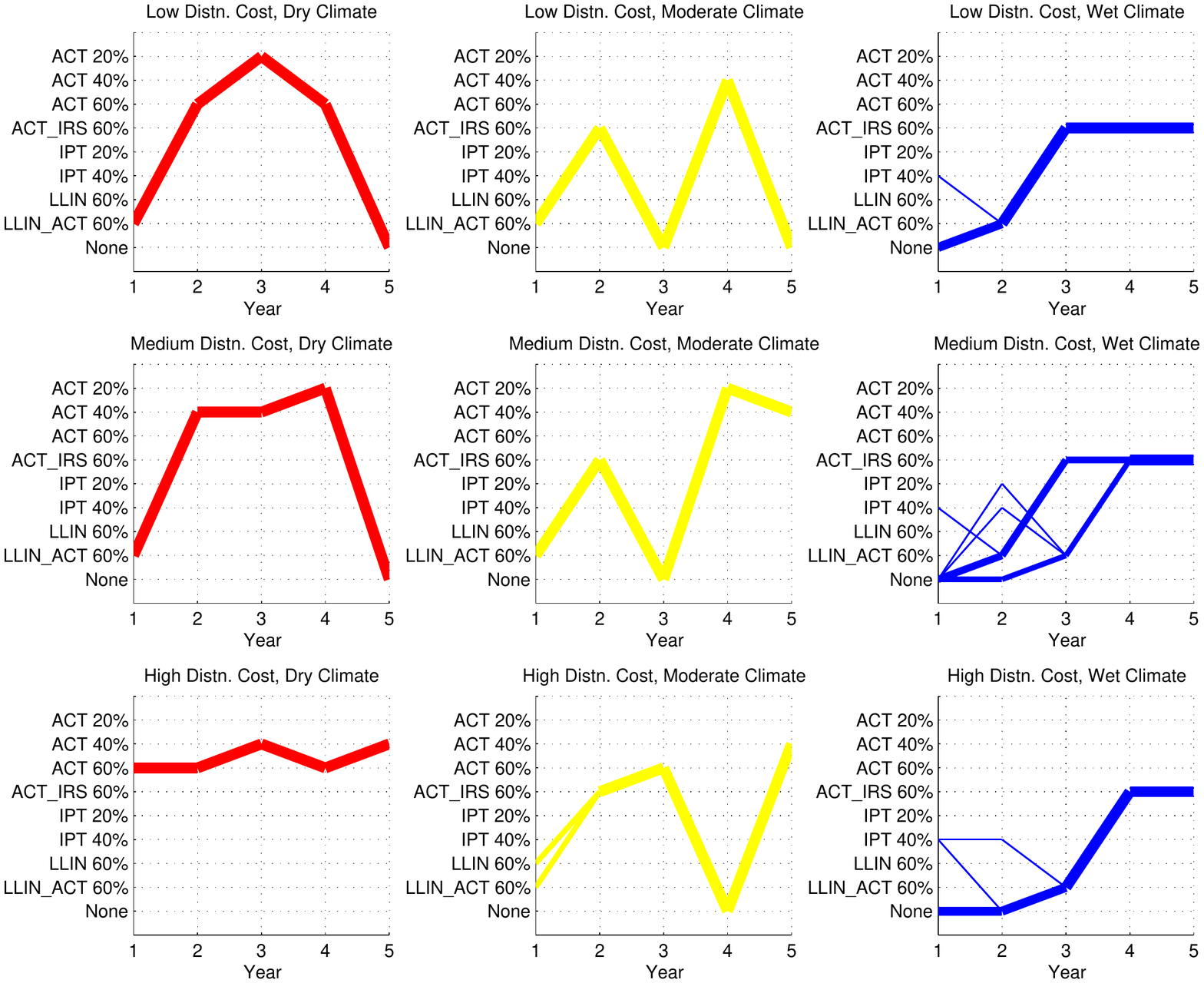}
    \caption{Sequence of interventions over a five-year time horizon in the optimistic efficacy scenario.  The thickness of each line is proportional to the number of districts assigned a given sequence.  Intervention sequences assigned to only a single district have been omitted from the figure (4 districts omitted in total).  The total person-days of malaria infection in this scenario is 2.977 billion.}
    \label{fig:optimisticoutput}
    \end{figure}

\begin{figure}[h!]
\includegraphics[width=\textwidth]{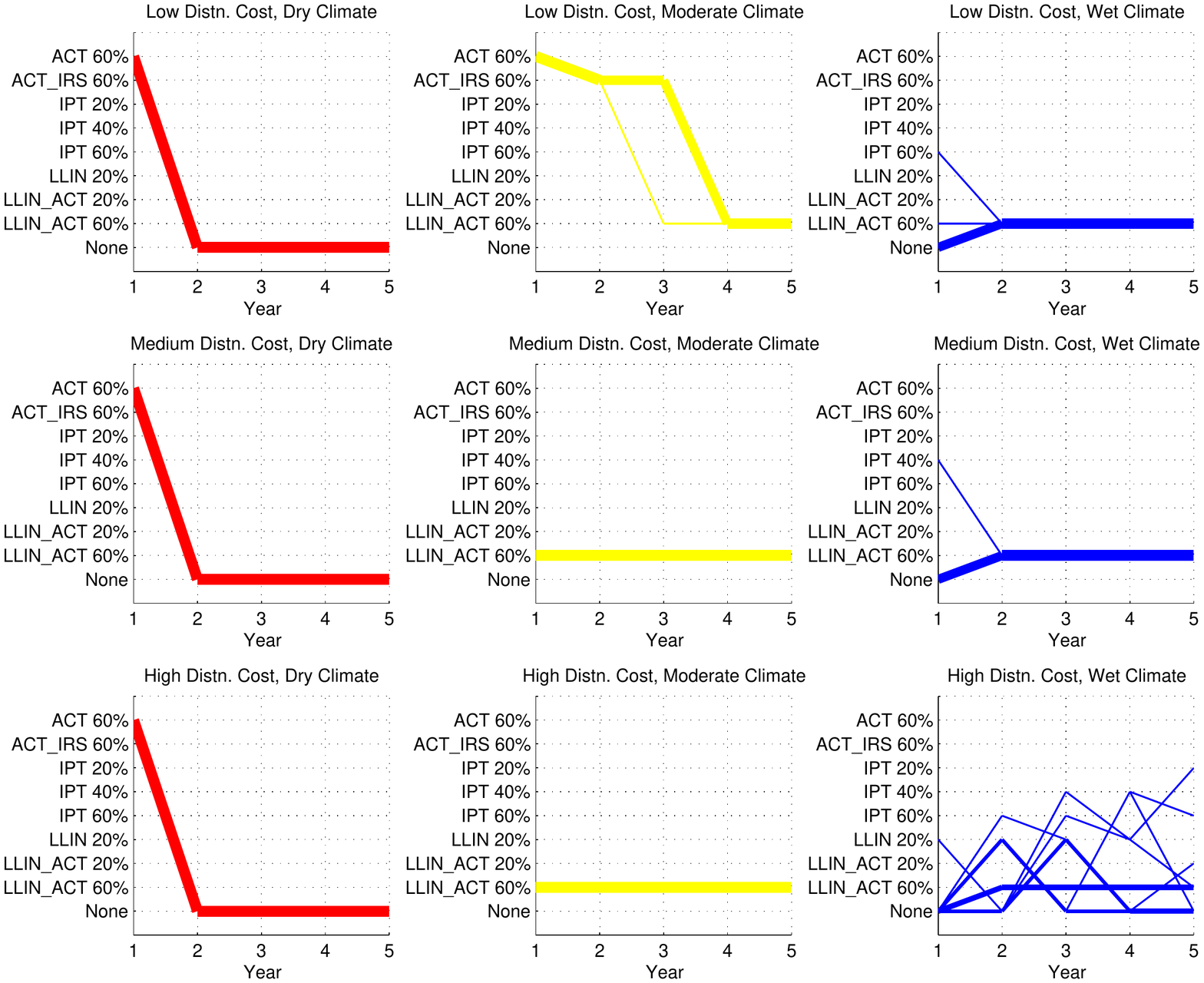}
    \caption{Sequence of interventions over a five-year time horizon in the pessimistic efficacy scenario.  The thickness of each line is proportional to the number of districts assigned a given sequence.  Intervention sequences assigned to only a single district have been omitted from the figure (5 districts omitted in total).  The total person-days of malaria infection in this scenario is 5.080 billion.}
    \label{fig:pessimisticoutput}
    \end{figure}

We can examine the sequence of interventions allocated in the nine geographic regions under our optimistic and pessimistic efficacy scenarios, in which the disease transmission parameter values for the treated class are adjusted up or down by 30\%.  The optimistic case is depicted in Figure \ref{fig:optimisticoutput}, with the full set of results given in Table \ref{tab:optimisticoutput}, and the results for the pessimistic case are given in Figure \ref{fig:pessimisticoutput} and Table \ref{tab:pessimisticoutput}. 

In the optimistic case, we see a reduction in person-days of malaria infection from $4.506$ billion to $2.977$  billion, or 34\%.  In year one, the model focuses on eradicating (subject to rounding error) malaria in the dry and moderate climate regions by allocating LLIN, ACT or the two in combination.  All but seven of the 1500 wet climate districts receive no intervention in year one.  In subsequent years, the wet climate districts receive 60\% coverage of LLIN with ACT, and ACT with IRS, with a handful of wet climate districts receiving IPT or no intervention during years one and two.  Although it appears that the dry and moderate regions are also receiving interventions during years two through five, this is an artifact of the optimization model: once the infectious population is driven to zero, there is no value in distributing further interventions; the model is allocating interventions in these regions simply to use up the available budget.  Moreover, the prevalence of malaria in the wet regions appears to stabilize around 10\% in year five.  This is further indication that 60\% coverage is not sufficient to diminish malaria prevalence in the wet regions, even if we are optimistic about the interventions' efficacy.

In the pessimistic case, we see an increase in person-days of malaria infection from 4.506 billion to 5.080 billion, or 13\%.  In year one, resources are focused on rapidly reducing infections in the dry regions, by allocating 60\% coverage of ACT; remaining resources in year one are focused on the moderate climate regions.  In subsequent years, the moderate regions receive 60\% coverage of ACT combined with either IRS or LLIN; the remaining budget is used to allocate a variety of interventions in the wet regions.  The takeaway message here is that to minimize person-days of malaria infection, the model chooses to focus resources on the dry and moderate climate regions; any remaining budget is allocated to the wet regions. 

\subsection*{Effect of Coverage}
If we assume that a maximum coverage of 80\% (including both distribution and compliance) is attainable, we see that the results change dramatically.  Table \ref{tab:baselinehighcoverage} gives the sequence of interventions assuming a baseline efficacy and possible coverage levels of 40\%, 60\% and 80\% (rather than 20\%, 40\% and 60\% used earlier), and using the same annual budget of 33.75 million USD.  We see that in all  geographic regions, the infected population is driven to zero within two years, indicating that malaria is effectively eradicated (at least subject to rounding at our resolution of 5\%).  Thus, achieving high coverage is crucial to rapid eradication of malaria, even with the annual budget held constant. Although our intervention costs rise in proportion to the treatment coverage, our fixed budget is able to achieve markedly lower person-days of malaria infection when 80\% coverage is achievable as opposed to only 60\% (1.139 billion person days, as opposed to 4.506 billion person days,) although some of this effect is likely due to rounding at resolution 5\%.   Thus, giving a smaller number of cities a higher coverage is more effective than giving a larger number of cities a lower coverage.  Moreover, it is only when 80\% coverage is attainable that we see a reduction of malaria prevalence in the wet climate regions.

Because the 80\% coverage eradicates malaria so quickly in our model, the interpretation of the five-year model results becomes less interesting.  For this reason, we present the remainder of the results using the original coverage percentages of 20\%, 40\% and 60\%,  except where otherwise indicated.

\subsection*{Effect of Budget}
\begin{figure}[h!]
\includegraphics[width=\textwidth]{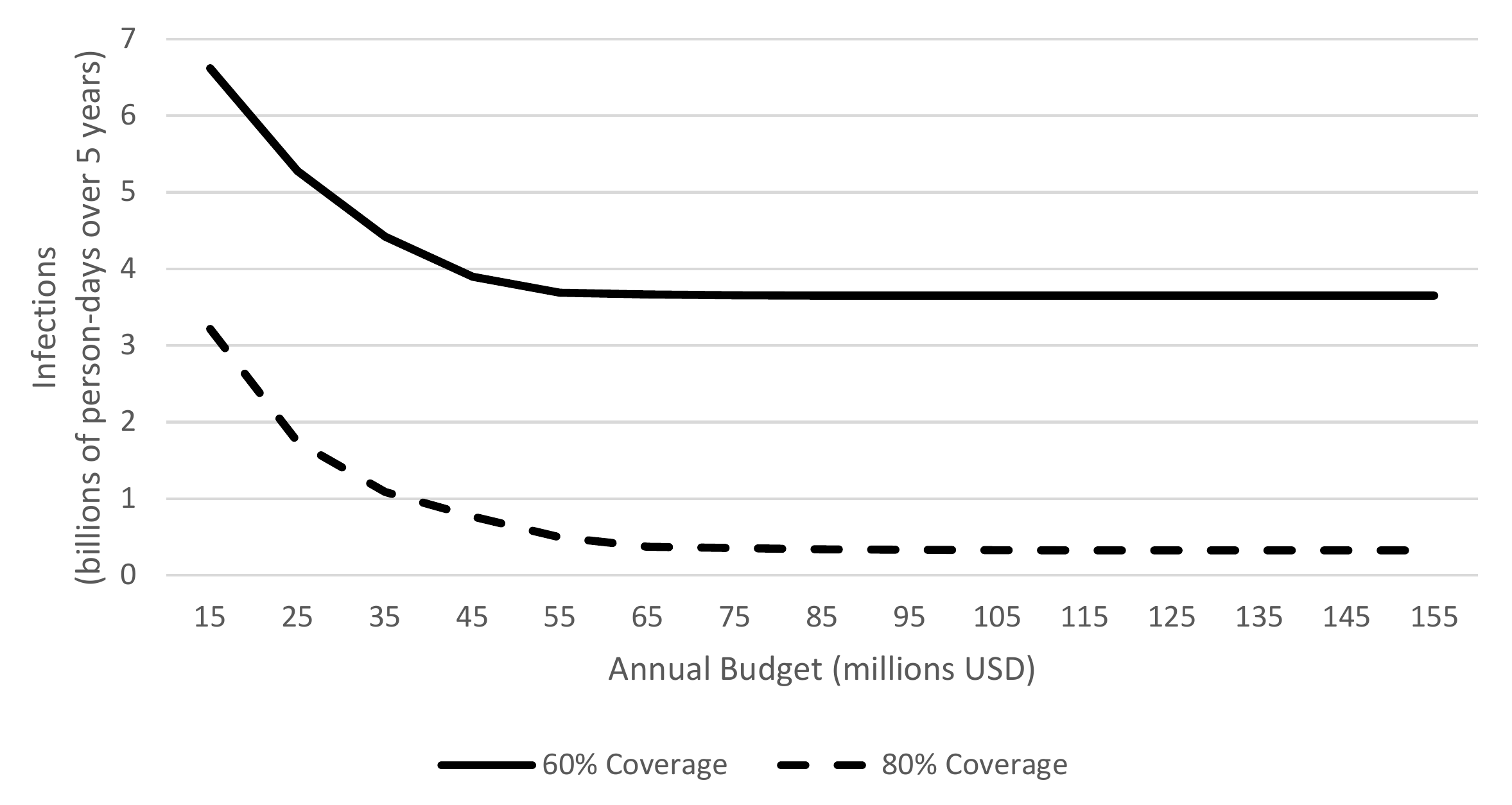}
\caption{Total person-days of malaria infection over five-year horizon as a function of annual budget in the baseline efficacy scenario. The solid line corresponds to a maximum attainable coverages of 60\%. The dashed line corresponds to a maximum attainable coverage of 80\%.}
    \label{fig:budVsIppl}
    \end{figure}

Previous work by  Dimitrov \textit{et al.} suggested that increasing the budget would decrease malaria deaths roughly linearly up to a critical budget value, after which there would be diminishing marginal benefit to additional budget expenditures \cite{dimitrov}.  Using the baseline efficacy scenario, we solved the optimization problem sequentially for budgets ranging from 15 million USD to  155 million USD in increments of 10 million USD and plotted person-days of malaria infection against budget.  As seen in Figure \ref{fig:budVsIppl}, our results are consistent with Dimitrov \textit{et al.} \cite{dimitrov}. We find that the person-days of malaria infection decrease linearly with an increase in budget until roughly 55 million USD, after which we observe diminishing marginal returns on budget increases.  Moreover, beyond a budget of 85 million USD, we observe no further reduction in person-days of malaria infection.  Also shown in Figure \ref{fig:budVsIppl} is the plot of person-days of malaria infection in the baseline scenario when coverage up to 80\% is attainable.  We see the same general shape to the graph, but with substantially lower values for the person-days of malaria infection.  Moreover, we see that even with a very large budget, 60\% coverage cannot achieve the low prevalence of malaria infection that 80\% coverage can achieve with even a small budget.   

\begin{figure}[h!]
\includegraphics[width=\textwidth]{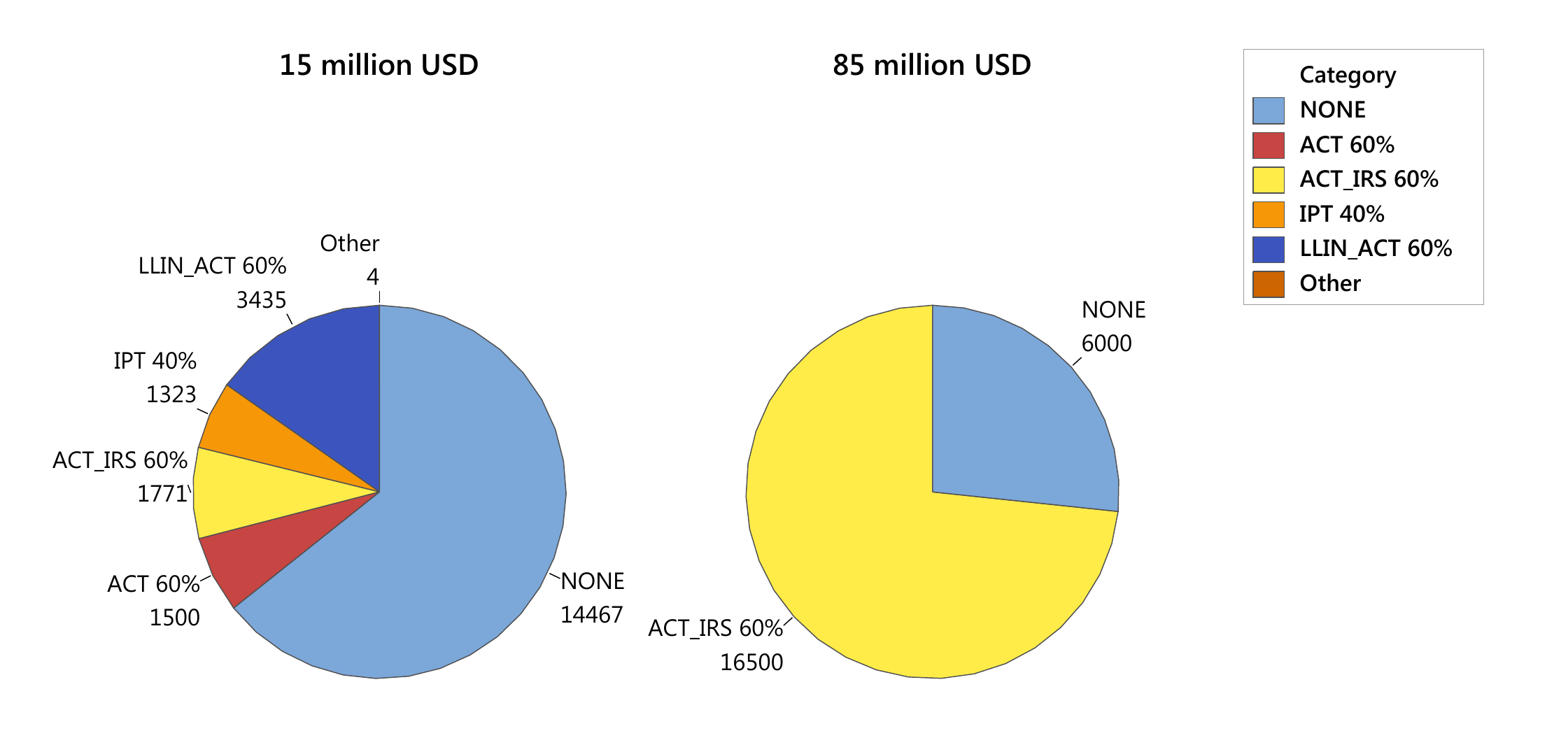}
\caption{Interventions chosen (in district-years) under a 15 million USD annual budget as compared to an 85 million USD annual budget, aggregated over all geographic regions and a five-year time horizon in the baseline efficacy scenario.}
    \label{fig:budgets}
    \end{figure}

Additionally, we can examine the qualitative change in the interventions chosen at different budget levels, as shown in Figure \ref{fig:budgets}.  When the budget is 15 million USD, for instance, most districts receive no intervention in most years.  Those that do receive 60\% coverage of ACT, either alone or in combination with either IRS or LLIN, or 40\% coverage of IPT.  (Although not shown in Figure \ref{fig:budgets}, the raw results from the simulation reveal that the dry regions receive interventions in year one, which effectively eradicates malaria in those regions; most of the moderate climate regions  receive interventions in most years, with a cluster of high distribution cost districts receiving no intervention in any of the five years; and the wet regions typically receive no intervention in most years.)  When the budget is 85 million USD, Figure \ref{fig:budgets} shows that the number of districts receiving no intervention in some year drops sharply.  (A closer look at the raw results reveals that all regions receive the maximum of 60\% coverage of ACT in combination with IRS in all years, except for the dry regions, which receive no intervention in years two through five after malaria has been effectively eradicated.)

\begin{figure}[h!]
\includegraphics[width=\textwidth]{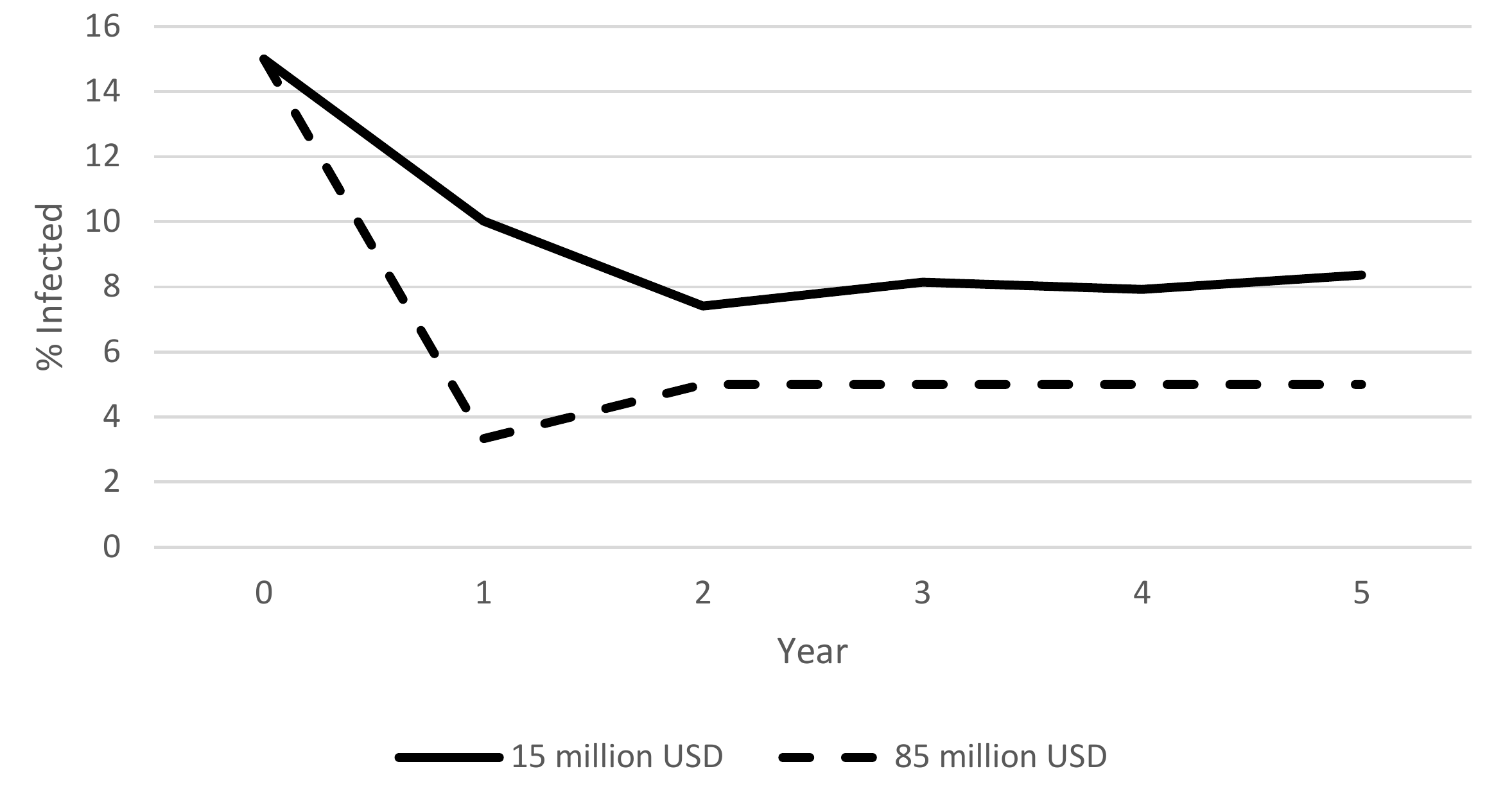}
\caption{Percentage of the population in the infected class over time under a 15 million USD annual budget (solid line)  as compared to an 85 million USD annual budget (dashed line), aggregated over all geographic regions, in the baseline efficacy scenario.}
\label{fig:percentinfected}
\end{figure}

Figure \ref{fig:percentinfected} shows the percentage of the total population that is in the infected class at the end of each of the five years, when the annual budget is 15 million USD as compared to 85 million USD.  A 15 million USD annual budget achieves a drop from 15\% infected to around 8\% infected in steady state.  An 85 million USD annual budget achieves an initial drop to around 3\% infected, which levels out to 5\% infected  in steady-state.  (Note that this contrasts with the results shown previously in Table \ref{tab:baselinehighcoverage} where if coverage of 80\% is attainable, malaria is eradicated with only a $33.75$ million USD budget.)  

\subsection*{Role of Vaccine}
\begin{figure}[h!]
	\includegraphics[width=\textwidth]{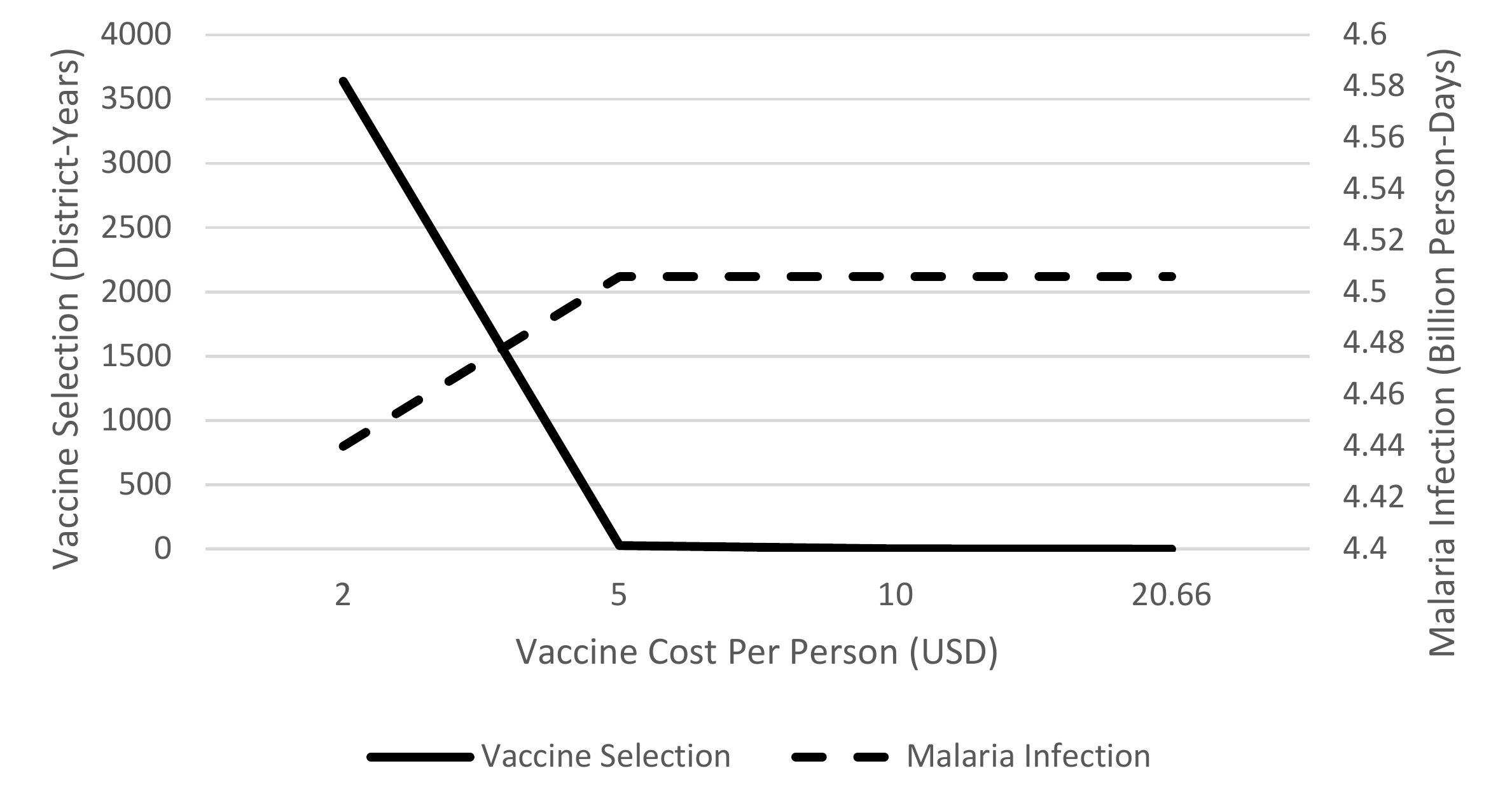}
	\caption{Vaccine selection (in number of district-years) and person-days of malaria infection versus vaccine cost, over 5-year time horizon, in the baseline efficacy scenario.}
    \label{fig:ipplDays_VaccineUse_vs_Cost}
    \end{figure}

We were surprised to observe that in all of our simulations across geographic regions, efficacy scenarios, maximum coverage, and budgets, the vaccine was almost never chosen, either alone or in combination with IPT.  Although the annual cost per treated person is quite high (20.66 USD), the fact that it is distributed only to children under the age of four, comprising an estimated 14.6\% of the population, makes its district-wide cost on par with that of the other interventions.  We can use our model to understand under what conditions  a malaria vaccine would be a cost-effective intervention by systematically lowering its cost and increasing its efficacy.  Specifically, we solved the optimization model on all combinations of $b_t = 0.001, 0.005,$ or $0.009$, $\gamma_t = \frac{1}{2}, \frac{1}{5.5},$ or  $\frac{1}{30}$, and vaccine cost $ = 2, 5, 10,$ or $20.66$ USD per person annually.  Preliminary analysis of these results showed little influence of either $b_t$ or $\gamma_t$ on the number of times vaccine was selected; for most combinations of $b_t$ and $\gamma_t$, vaccine will be frequently chosen if the cost is $2$ or $5$ USD and will rarely be chosen if the cost is $10$ or $20.66$ USD.  Therefore, we conclude that cost is a driving factor in the choice to select vaccines over other interventions.  Figure \ref{fig:ipplDays_VaccineUse_vs_Cost} shows the number of districts receiving vaccine, either alone or in combination with IPT, over the 5-year time horizon as a function of vaccine cost, for the baseline efficacy scenario and using $b_t=0.005,$ and $\gamma_t=\frac{1}{5.5}$.  We see that when the vaccine cost is $2$ USD per person, it is selected quite often, but is selected far less frequently when the cost is $5$ USD.  Moreover, it is almost never selected at a cost of $10$ or $20.66$.  Shown in the same figure is the person-days of malaria infection in those same instances.  We see that reducing the vaccine cost from $20.66$ USD per person to $2$ USD per person yields only a 1.5\% reduction in person-days of malaria infection, from $4.51$ billion to $4.44$ billion.    

\subsection*{Role of Time Horizon}
\begin{figure}[h!]
\includegraphics[width=\textwidth]{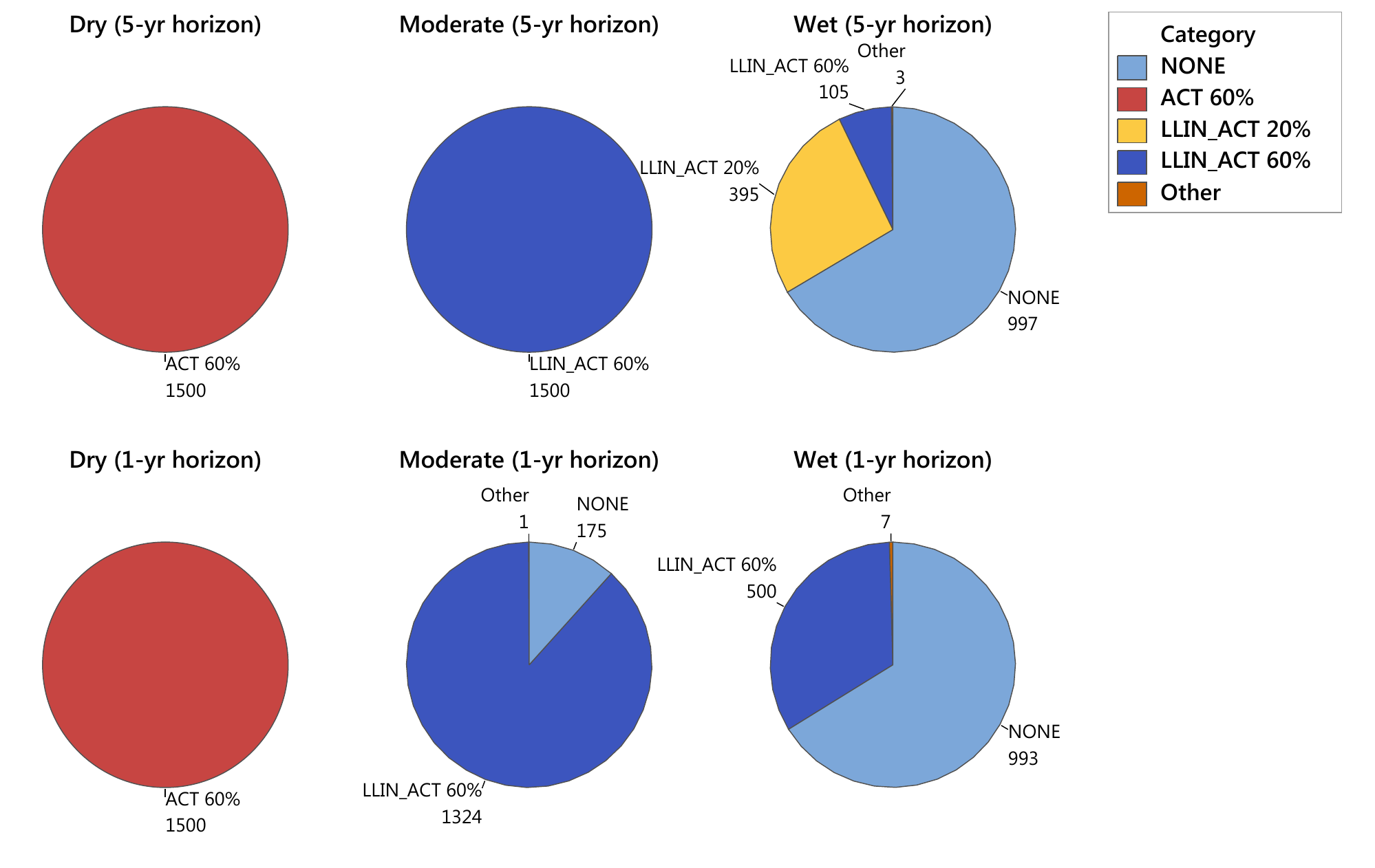}
	\caption{First-year interventions chosen in a one-year time horizon versus a five-year time horizon, by climate region, in the baseline efficacy scenario.}
    \label{fig:oneyear}
    \end{figure}

An important motivation underlying this work is the idea that distributing an intervention changes the malaria transmission dynamics, which could affect future intervention choices.  We can compare the interventions chosen in the first year of our five-year time horizon to those chosen if we consider a time horizon of only a single year, as shown in Figure \ref{fig:oneyear}.  As mentioned above, in the first year of the five-year time horizon, resources are focused primarily on the dry and moderate climate regions that receive either ACT alone at 60\% coverage, or ACT in combination with LLIN at 60\% coverage.  Most of the medium-to-high distribution cost districts in the wet regions receive no treatment during year one.  Nearly 400 of the low distribution cost districts in the wet regions get 20\% coverage of LLIN with ACT, and the remaining 100 get 60\% coverage of LLIN with ACT.  By contrast, if we solve the model using only a one-year time horizon, the solution changes.  Dry regions (at all distribution costs) again receive ACT alone at 60\% coverage. Moderate climate regions having low or medium distribution costs receive LLIN in combination with ACT at 60\% coverage, as do 324 moderate climate regions having high distribution costs.  However, now 175 moderate climate districts having high distribution costs drop down to no intervention, and in return, all 500 wet climate districts  having low distribution costs receive LLIN with ACT at 60\% coverage, which is substantially higher than before. Thus, we see that incorporating a multi-year planning horizon that anticipates changes in the malaria transmission as a result of the interventions distributed does indeed affect the optimal choice of intervention.

\subsection*{Role of Climate}
\begin{figure}[h!]
\includegraphics[width=\textwidth]{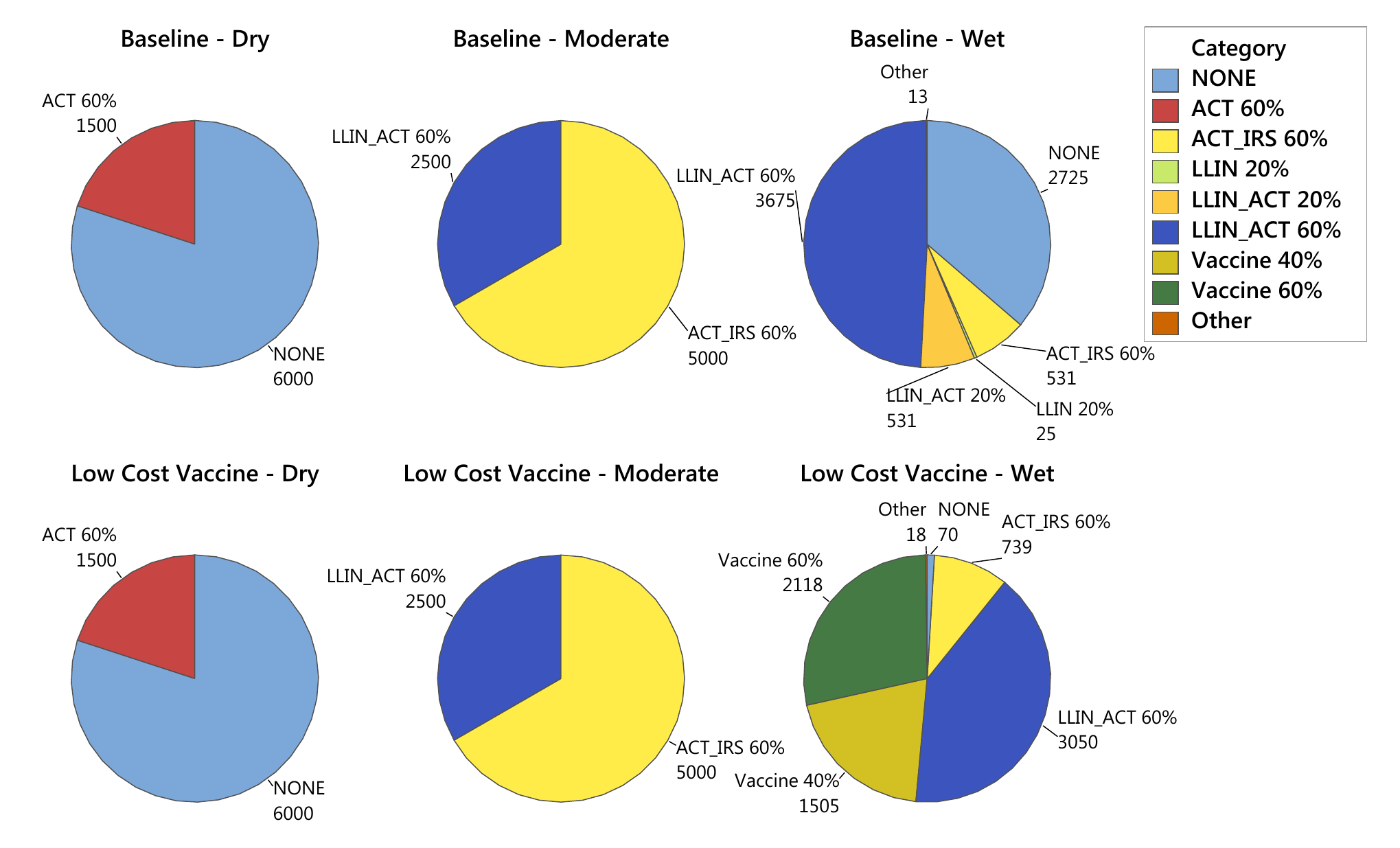}
\caption{Interventions chosen by climate region when vaccine cost is 20.66 USD per treated person versus a low-cost vaccine having cost 2 USD per treated person, in the baseline efficacy scenario.}
\label{fig:climate}
\end{figure}

There is a noticeable qualitative difference in the types of treatments distributed to the three climate regions considered.  Aggregating over all years and distribution cost categories for the baseline efficacy scenario, the first row of Figure \ref{fig:climate} shows the number of districts receiving each type of treatment, separated by climate region.  We see that dry regions are adequately served by ACT alone; moderate regions require ACT in combination with either LLIN or IRS; and wet regions get LLIN alone, LLIN with ACT, ACT with IRS, and a handful of others (though as discussed earlier, this variety is likely the result of the model using the available budget because no treatment at 60\% coverage is very successful at reducing malaria in the wet regions.)

If a lower-cost vaccine is available at $2$ USD per person per year, the second row of Figure \ref{fig:climate} shows that dry regions again receive exclusively no intervention and ACT, and moderate regions again receive exclusively ACT in combination with either IRS or LLIN. Now, however, in the wet regions, the number of years of ``no intervention'' drops significantly, from 2725 to 70, and we see that vaccine use, at either  40\% or 60\% coverage comprises nearly 50\% of the pie chart.   

The takeaway messages are that a single choice of intervention across all climate types is unlikely to be optimal, and that the development of a low-cost malaria vaccine is likely to be of greatest use in regions with high mosquito densities; drier areas are better served by ACT, alone or in combination with IRS and LLIN. 

\section*{Conclusions}

Because our model is very sensitive to the disease transmission parameter values used, it is best suited for qualitative interpretations about relative benefits of certain interventions.  For instance, while common sense might suggest we should focus intervention resources on wet climate regions with high mosquito counts, the results of our model with a maximum coverage of 60\% suggest the opposite: to reduce person-days of malaria infection, it might be better to invest resources on areas where interventions can dramatically reduce the prevalence of malaria, rather than expend resources on areas where malaria is likely to persist, despite our best efforts.  But if coverage closer to 80\% is attainable, then malaria can be combatted in wet climate regions.  Likewise, our model can illustrate the sensitivity of the optimal policy to certain parameters.  For example, the optimal sequence of interventions varies by climate type (as represented by mosquito density), suggesting that a one-size-fits-all approach to malaria eradication is not optimal.  The sensitivity of the model to parameter assumptions also signals that prior to using the model to guide policy in any given region, the choice of parameter values would first need to be calibrated to match known malaria prevalence rates in the region.


Future refinements of the model could address acquired resistance to interventions, age-dependent immunity, spatial effects of human or mosquito migration, and computational tractability.  Drug resistance in malaria parasites and insecticide resistance in mosquitoes are major challenges to control and eradication efforts \cite{WHO-drugresistance, kim+schneider2013, edi2012}.  Implementing resistance in this model would require tracking decreased effectiveness of treatment after use in multiple consecutive years.  This would increase the computational challenge of solving the model, however, as the costs and benefits of choosing a particular intervention in a given year would depend not only on the $(S,I,R)$ population state but also on the sequence of interventions chosen in prior years. Likewise, incorporating age-dependent immunity or spatial effects would also require an increase in the number of population compartments in the $(S,I,R)$ model. Already, we were limited by computational power in the resolution of the population state we could consider.  The bottleneck appears to be in the formulation of the integer linear programming model using AMPL.  Although the data file is only 2MB at  5\% resolution and 11MB at  2\% resolution, loading the 5\% resolution data file into AMPL took approximately 15 minutes on a 32-core, 128 GB RAM parallel compute server located in the Harvey Mudd College Mathematics Department, and attempting to load the data file for the 2\% resolution case exceeded the available 128 GB of RAM on the server.  Once loaded, the 5\% resolution model was subsequently solved by the CPLEX solver within seconds.   The technical staff at the NEOS server (an online server for optimization solvers on which we ran earlier tests \cite{czyzyk+all1998, dolan2001, gropp+more1997}) who are familiar with these modeling languages suggested using a more efficient modeling language  than AMPL for formulating the optimization model; we leave this as future work.

Given the growing interest in malaria eradication, the WHO Global Malaria Programme cites the need for operations research studies to determine the best intervention strategies in areas where transmission dynamics are changing as malaria is being eliminated.  They also present a list of priority research questions that includes questions about safety, access, and community involvement \cite{planningWHO-2013}.  We have developed a flexible modeling framework that can guide such decisions.  Our model permits a multi-year planning horizon over areas characterized by disparate infrastructure and climate. Given inputs of the per-person cost of each intervention and the effects each intervention has on malaria disease transmission parameters, the model provides a sequence of interventions over a fixed time horizon that minimizes person-days of malaria infection subject to an annual budget.  Moreover, this model can be adapted to the treatment of other infectious diseases.


\section*{Competing interests}
  The authors declare that they have no competing interests.

\section*{Author's contributions}
    HJD and SEM wrote the paper.  HJD, AG, and CJO conducted literature review, gathered data, implemented the computational model, wrote the simulations, and wrote an initial draft of the paper.  SEM proposed the concept for the model, guided model development and analysis, finalized the computational model and simulations, analyzed the model, and secured funding for the work.  

\section*{Acknowledgements}
This project was supported by grants to Harvey Mudd College for student and faculty research from the Arnold and Mabel Beckman Foundation, the Howard Hughes Medical Institute, the Fletcher Jones Foundation, the Andrew W. Mellon Foundation, and The Rose Hills Foundation.  Additionally, we thank Shanika Lazo for her contributions to this work, and Katarina Hoeger, Tracey Luke, Taryn Ohashi, Tristan Williams and Flora Xu for their contributions to an earlier version of the model presented here.  We also thank two anonymous reviewers for their suggestions which greatly improved this work.



\bibliographystyle{bmc-mathphys}

\bibliography{Malaria_arxiv}


\begin{thebibliography}{46}
\ifx \bisbn   \undefined \def \bisbn  #1{ISBN #1}\fi
\ifx \binits  \undefined \def \binits#1{#1}\fi
\ifx \bauthor  \undefined \def \bauthor#1{#1}\fi
\ifx \batitle  \undefined \def \batitle#1{#1}\fi
\ifx \bjtitle  \undefined \def \bjtitle#1{#1}\fi
\ifx \bvolume  \undefined \def \bvolume#1{\textbf{#1}}\fi
\ifx \byear  \undefined \def \byear#1{#1}\fi
\ifx \bissue  \undefined \def \bissue#1{#1}\fi
\ifx \bfpage  \undefined \def \bfpage#1{#1}\fi
\ifx \blpage  \undefined \def \blpage #1{#1}\fi
\ifx \burl  \undefined \def \burl#1{\textsf{#1}}\fi
\ifx \doiurl  \undefined \def \doiurl#1{\textsf{#1}}\fi
\ifx \betal  \undefined \def \betal{\textit{et al.}}\fi
\ifx \binstitute  \undefined \def \binstitute#1{#1}\fi
\ifx \binstitutionaled  \undefined \def \binstitutionaled#1{#1}\fi
\ifx \bctitle  \undefined \def \bctitle#1{#1}\fi
\ifx \beditor  \undefined \def \beditor#1{#1}\fi
\ifx \bpublisher  \undefined \def \bpublisher#1{#1}\fi
\ifx \bbtitle  \undefined \def \bbtitle#1{#1}\fi
\ifx \bedition  \undefined \def \bedition#1{#1}\fi
\ifx \bseriesno  \undefined \def \bseriesno#1{#1}\fi
\ifx \blocation  \undefined \def \blocation#1{#1}\fi
\ifx \bsertitle  \undefined \def \bsertitle#1{#1}\fi
\ifx \bsnm \undefined \def \bsnm#1{#1}\fi
\ifx \bsuffix \undefined \def \bsuffix#1{#1}\fi
\ifx \bparticle \undefined \def \bparticle#1{#1}\fi
\ifx \barticle \undefined \def \barticle#1{#1}\fi
\ifx \bconfdate \undefined \def \bconfdate #1{#1}\fi
\ifx \botherref \undefined \def \botherref #1{#1}\fi
\ifx \url \undefined \def \url#1{\textsf{#1}}\fi
\ifx \bchapter \undefined \def \bchapter#1{#1}\fi
\ifx \bbook \undefined \def \bbook#1{#1}\fi
\ifx \bcomment \undefined \def \bcomment#1{#1}\fi
\ifx \oauthor \undefined \def \oauthor#1{#1}\fi
\ifx \citeauthoryear \undefined \def \citeauthoryear#1{#1}\fi
\ifx \endbibitem  \undefined \def \endbibitem {}\fi
\ifx \bconflocation  \undefined \def \bconflocation#1{#1}\fi
\ifx \arxivurl  \undefined \def \arxivurl#1{\textsf{#1}}\fi
\csname PreBibitemsHook\endcsname

\bibitem{WHO2014}
\begin{botherref}
\oauthor{\bsnm{{World Health Organization}}}:
Malaria Fact Sheet.
Online.
http://www.who.int/mediacentre/factsheets/fs094/en/
(Accessed {M}arch 2014)
\end{botherref}
\endbibitem

\bibitem{planningWHO-2013}
\begin{botherref}
\oauthor{\bsnm{{World Health Organization: Global Malaria Programme}}}:
Meeting report: Planning meeting for operational research on malaria
  elimination.
Technical report
(October 2013)
\end{botherref}
\endbibitem

\bibitem{CDC2014}
\begin{botherref}
\oauthor{\bsnm{{Centers for Disease Control and Prevention}}}:
Insecticide-treated bed nets.
Online.
http://www.cdc.gov/malaria/malaria\_worldwide/reduction/itn.html.
(Accessed {M}arch 2014)
\end{botherref}
\endbibitem

\bibitem{whopes2005}
\begin{botherref}
{World Health Organization Pesticide Evaluation Scheme (WHOPES)}:
Guidelines for Laboratory and Field Testing of Long-lasting Insecticidal
  Mosquito Nets.
(2013).
{World Health Organization Pesticide Evaluation Scheme (WHOPES)}. ISBN 978 92 4
  150527 7
\end{botherref}
\endbibitem

\bibitem{pluess2010}
\begin{botherref}
\oauthor{\bsnm{Pluess}, \binits{B.}},
\oauthor{\bsnm{Tanser}, \binits{F.C.}},
\oauthor{\bsnm{Lengeler}, \binits{C.}},
\oauthor{\bsnm{Sharpe}, \binits{B.L.}}:
Indoor residual spraying for preventing malaria (review).
The Cochrane Library
(2010).
doi: 10.1002/14651858.CD006657.pub2
\end{botherref}
\endbibitem

\bibitem{aponte2009}
\begin{barticle}
\bauthor{\bsnm{Aponte}, \binits{J.J.}},
\bauthor{\bsnm{Schellenberg}, \binits{D.}},
\bauthor{\bsnm{Egan}, \binits{A.}},
\bauthor{\bsnm{Breckenridge}, \binits{A.}},
\bauthor{\bsnm{Carneiro}, \binits{I.}},
\bauthor{\bsnm{Critchley}, \binits{J.}},
\bauthor{\bsnm{Danquah}, \binits{I.}},
\bauthor{\bsnm{Dodoo}, \binits{A.}},
\bauthor{\bsnm{Kobbe}, \binits{R.}},
\bauthor{\bsnm{Lell}, \binits{B.}},
\bauthor{\bsnm{May}, \binits{J.}},
\bauthor{\bsnm{Premji}, \binits{Z.}},
\bauthor{\bsnm{Sanz}, \binits{S.}},
\bauthor{\bsnm{Sevene}, \binits{E.}},
\bauthor{\bsnm{Soulamani-Becheikh}, \binits{R.}},
\bauthor{\bsnm{Winstanley}, \binits{P.}},
\bauthor{\bsnm{Adjei}, \binits{S.}},
\bauthor{\bsnm{Anemana}, \binits{S.}},
\bauthor{\bsnm{Chandramohan}, \binits{D.}},
\bauthor{\bsnm{Issifou}, \binits{S.}},
\bauthor{\bsnm{Mockenhaupt}, \binits{F.}},
\bauthor{\bsnm{Owusu-Agyei}, \binits{S.}},
\bauthor{\bsnm{Greenwood}, \binits{B.}},
\bauthor{\bsnm{Grobusch}, \binits{M.P.}},
\bauthor{\bsnm{Kremsner}, \binits{P.G.}},
\bauthor{\bsnm{Macete}, \binits{E.}},
\bauthor{\bsnm{Mshinda}, \binits{H.}},
\bauthor{\bsnm{Newman}, \binits{R.D.}},
\bauthor{\bsnm{Slutsker}, \binits{L.}},
\bauthor{\bsnm{Tanner}, \binits{M.}},
\bauthor{\bsnm{Alonso}, \binits{P.}},
\bauthor{\bsnm{Menendez}, \binits{C.}}:
\batitle{Efficacy and safety of intermittent preventive treatment with
  sulfadoxine-pyrimethamine for malaria in {A}frican infants: A pooled analysis
  of six randomized, placebo-controlled trials}.
\bjtitle{Lancet}
\bvolume{374}(\bissue{9700}),
\bfpage{1533}--\blpage{1542}
(\byear{2009}).
\bcomment{doi: 10.1016/S0140-6736(09)61258-7}
\end{barticle}
\endbibitem

\bibitem{doolan2009}
\begin{barticle}
\bauthor{\bsnm{Doolan}, \binits{D.L.}},
\bauthor{\bsnm{Doba\~{n}o}, \binits{C.}},
\bauthor{\bsnm{Baird}, \binits{J.K.}}:
\batitle{Acquired immunity to malaria}.
\bjtitle{Clinical Microbiology Reviews}
\bvolume{22}(\bissue{1}),
\bfpage{13}--\blpage{36}
(\byear{2009}).
\bcomment{doi: 10.1128/CMR.00025-08.}
\end{barticle}
\endbibitem

\bibitem{MC2014}
\begin{botherref}
\oauthor{\bsnm{{Malaria Consortium}}}:
Artemisinin-based combination therapy.
Online.
http://www.malariaconsortium.org/page.php?id=112.
(Accessed {A}ugust 2014)
\end{botherref}
\endbibitem

\bibitem{duffy2006}
\begin{barticle}
\bauthor{\bsnm{Duffy}, \binits{P.E.}},
\bauthor{\bsnm{Mutabingwa}, \binits{T.K.}}:
\batitle{Artemisinin combination therapies}.
\bjtitle{Lancet}
\bvolume{367}(\bissue{9528}),
\bfpage{2037}--\blpage{2039}
(\byear{2006}).
\bcomment{doi:10.1016/S0140-6736(06)68900-9}
\end{barticle}
\endbibitem

\bibitem{crompton+pierce+miller2010}
\begin{barticle}
\bauthor{\bsnm{Crompton}, \binits{P.D.}},
\bauthor{\bsnm{Pierce}, \binits{S.K.}},
\bauthor{\bsnm{Miller}, \binits{L.H.}}:
\batitle{Advances and challenges in malaria vaccine development}.
\bjtitle{The Journal of Clinical Investigation}
\bvolume{120}(\bissue{12}),
\bfpage{4168}--\blpage{4178}
(\byear{2010}).
\bcomment{doi:10.1172/JCI44423}
\end{barticle}
\endbibitem

\bibitem{dimitrov}
\begin{barticle}
\bauthor{\bsnm{Dimitrov}, \binits{N.B.}},
\bauthor{\bsnm{Moffett}, \binits{A.}},
\bauthor{\bsnm{Morton}, \binits{D.P.}},
\bauthor{\bsnm{Sarkar}, \binits{S.}}:
\batitle{Selecting malaria interventions: A top-down approach}.
\bjtitle{Computers \& Operations Research}
\bvolume{40}(\bissue{9}),
\bfpage{2229}--\blpage{2240}
(\byear{2013}).
\bcomment{doi: 10.1016/j.cor.2011.07.023}
\end{barticle}
\endbibitem

\bibitem{ross1915}
\begin{barticle}
\bauthor{\bsnm{Ross}, \binits{R.}}:
\batitle{Some a priori pathometric equations}.
\bjtitle{The British Medical Journal}
\bvolume{1}(\bissue{2830}),
\bfpage{546}--\blpage{547}
(\byear{1915}).
\bcomment{http://www.ncbi.nlm.nih.gov/pmc/articles/PMC2302055/pdf/brmedj07224-0010.pdf}
\end{barticle}
\endbibitem

\bibitem{mandal+sarkar+sinha2011}
\begin{botherref}
\oauthor{\bsnm{Mandal}, \binits{S.}},
\oauthor{\bsnm{Sarkar}, \binits{R.R.}},
\oauthor{\bsnm{Sinha}, \binits{S.}}:
Mathematical models of malaria - a review.
Malaria Journal
\textbf{10}(202)
(2011).
http://www.malariajournal.com/content/10/1/202
\end{botherref}
\endbibitem

\bibitem{lindblade+all2005}
\begin{barticle}
\bauthor{\bsnm{Lindblade}, \binits{K.A.}},
\bauthor{\bsnm{Dotson}, \binits{E.}},
\bauthor{\bsnm{Hawley}, \binits{W.A.}},
\bauthor{\bsnm{Bayoh}, \binits{N.}},
\bauthor{\bsnm{Williamson}, \binits{J.}},
\bauthor{\bsnm{Mount}, \binits{D.}},
\bauthor{\bsnm{Olang}, \binits{G.}},
\bauthor{\bsnm{Vulule}, \binits{J.}},
\bauthor{\bsnm{Slutsker}, \binits{L.}},
\bauthor{\bsnm{Gimnig}, \binits{J.}}:
\batitle{Evaluation of long-lasting insecticidal nets after 2 years of
  household use}.
\bjtitle{Tropical Medicine and International Health}
\bvolume{10}(\bissue{11}),
\bfpage{1141}--\blpage{1150}
(\byear{2005}).
\bcomment{doi:10.1111/j.1365-3156.2005.01501.x}
\end{barticle}
\endbibitem

\bibitem{killeen+all2006}
\begin{botherref}
\oauthor{\bsnm{Killeen}, \binits{G.F.}},
\oauthor{\bsnm{Kihonda}, \binits{J.}},
\oauthor{\bsnm{Lyimo}, \binits{E.}},
\oauthor{\bsnm{Oketch}, \binits{F.R.}},
\oauthor{\bsnm{Kotas}, \binits{M.E.}},
\oauthor{\bsnm{Mathenge}, \binits{E.}},
\oauthor{\bsnm{Schellenberg}, \binits{J.A.}},
\oauthor{\bsnm{Lengeler}, \binits{C.}},
\oauthor{\bsnm{Smith}, \binits{T.A.}},
\oauthor{\bsnm{Drakeley}, \binits{C.J.}}:
Quantifying behavioural interactions between humans and mosquitoes: Evaluating
  the protective efficacy of insecticidal nets against malaria transmission in
  rural {T}anzania.
BMC Infectious Diseases
\textbf{6}(161)
(2006).
doi:10.1186/1471-2334-6-161
\end{botherref}
\endbibitem

\bibitem{bousema+all2010}
\begin{botherref}
\oauthor{\bsnm{Bousema}, \binits{T.}},
\oauthor{\bsnm{Okell}, \binits{L.}},
\oauthor{\bsnm{Shekalaghe}, \binits{S.}},
\oauthor{\bsnm{Griifin}, \binits{J.T.}},
\oauthor{\bsnm{Omar}, \binits{S.}},
\oauthor{\bsnm{Sawa}, \binits{P.}},
\oauthor{\bsnm{Sutherland}, \binits{C.}},
\oauthor{\bsnm{Sauerwein}, \binits{R.}},
\oauthor{\bsnm{Ghani}, \binits{A.C.}},
\oauthor{\bsnm{Drakeley}, \binits{C.}}:
Research revisiting the circulation time of {Plasmodium falciparum}
  gametocytes: Molecular detection methods to estimate the duration of
  gametocyte carriage and the effect of gametocytocidal drugs.
Malaria Journal
\textbf{9}(136)
(2010).
http://www.malariajournal.com/content/9/1/136
\end{botherref}
\endbibitem

\bibitem{garner+graves2005}
\begin{botherref}
\oauthor{\bsnm{Garner}, \binits{P.}},
\oauthor{\bsnm{Graves}, \binits{P.M.}}:
The benefits of artemisinin combination therapy for malaria extend beyond the
  individual patient.
PLoS Medicine
\textbf{2}(4)
(2005).
doi:10.1371/journal.pmed.0020105
\end{botherref}
\endbibitem

\bibitem{chandrahoman2005}
\begin{botherref}
\oauthor{\bsnm{Chandrahoman}, \binits{D.}},
\oauthor{\bsnm{Owuso-Ageyi}, \binits{S.}},
\oauthor{\bsnm{Carneiro}, \binits{I.}},
\oauthor{\bsnm{Awine}, \binits{T.}},
\oauthor{\bsnm{Amponsa-Achiano}, \binits{K.}},
\oauthor{\bsnm{Mensah}, \binits{N.}},
\oauthor{\bsnm{Jaffar}, \binits{S.}},
\oauthor{\bsnm{Baiden}, \binits{R.}},
\oauthor{\bsnm{Hodgson}, \binits{A.}},
\oauthor{\bsnm{Binka}, \binits{F.}},
\oauthor{\bsnm{Greenwood}, \binits{B.}}:
Cluster randomised trial of intermittent preventive treatment for malaria in
  infants in area of high, seasonal transmission in {G}hana.
British Medical Journal
\textbf{331}(727)
(2005).
doi: http://dx.doi.org/10.1136/bmj.331.7519.727
\end{botherref}
\endbibitem

\bibitem{grobusch2007}
\begin{barticle}
\bauthor{\bsnm{Grobusch}, \binits{M.P.}},
\bauthor{\bsnm{Lell}, \binits{B.}},
\bauthor{\bsnm{Schwarz}, \binits{N.G.}},
\bauthor{\bsnm{Gabor}, \binits{J.}},
\bauthor{\bsnm{D{\"o}rnemann}, \binits{J.}},
\bauthor{\bsnm{P{\"o}tschke}, \binits{M.}},
\bauthor{\bsnm{Oyakhirome}, \binits{S.}},
\bauthor{\bsnm{Kiessling}, \binits{G.C.}},
\bauthor{\bsnm{Necek}, \binits{M.}},
\bauthor{\bsnm{L{\"a}ngin}, \binits{M.U.}},
\bauthor{\bsnm{Klouwenberg}, \binits{P.K.}},
\bauthor{\bsnm{Kl{\"o}pfer}, \binits{A.}},
\bauthor{\bsnm{Naumann}, \binits{B.}},
\bauthor{\bsnm{Altun}, \binits{H.}},
\bauthor{\bsnm{Agnandji}, \binits{S.T.}},
\bauthor{\bsnm{Goesch}, \binits{J.}},
\bauthor{\bsnm{Decker}, \binits{M.}},
\bauthor{\bsnm{Salazar}, \binits{C.L.O.}},
\bauthor{\bsnm{Supan}, \binits{C.}},
\bauthor{\bsnm{Kombila}, \binits{D.U.}},
\bauthor{\bsnm{Borchert}, \binits{L.}},
\bauthor{\bsnm{Ko{\"o}ter}, \binits{K.B.}},
\bauthor{\bsnm{Pongratz}, \binits{P.}},
\bauthor{\bsnm{Adegnika}, \binits{A.A.}},
\bauthor{\bsnm{{von Glasenapp}}, \binits{I.}},
\bauthor{\bsnm{Issifou}, \binits{S.}},
\bauthor{\bsnm{Kremsner}, \binits{P.G.}}:
\batitle{Intermittent preventive treatment against malaria in infants in
  {G}abon: A randomized, double-blind, placebo-controlled trial}.
\bjtitle{Journal of Infectious Diseases}
\bvolume{196}(\bissue{1}),
\bfpage{1595}--\blpage{1602}
(\byear{2007}).
\bcomment{doi: 10.1086/522160}
\end{barticle}
\endbibitem

\bibitem{prosper+more2014}
\begin{barticle}
\bauthor{\bsnm{Prosper}, \binits{O.}},
\bauthor{\bsnm{Ruktanonchai}, \binits{N.}},
\bauthor{\bsnm{Martcheva}, \binits{M.}}:
\batitle{Optimal vaccination and bed net maintenance for the control of malaria
  in a region with naturally acquired immunity}.
\bjtitle{Journal of Theoretical Biology}
\bvolume{353},
\bfpage{142}--\blpage{156}
(\byear{2014}).
\bcomment{doi: 10.1016/j.jtbi.2014.03.013}
\end{barticle}
\endbibitem

\bibitem{bojang2001}
\begin{barticle}
\bauthor{\bsnm{Bojang}, \binits{K.A.}},
\bauthor{\bsnm{Milligan}, \binits{P.J.}},
\bauthor{\bsnm{Pinder}, \binits{M.}},
\bauthor{\bsnm{Vigneron}, \binits{L.}},
\bauthor{\bsnm{Alloueche}, \binits{A.}},
\bauthor{\bsnm{Kester}, \binits{K.E.}},
\bauthor{\bsnm{Ballou}, \binits{W.R.}},
\bauthor{\bsnm{Conway}, \binits{D.J.}},
\bauthor{\bsnm{Reece}, \binits{W.H.}},
\bauthor{\bsnm{Gothard}, \binits{P.}},
\bauthor{\bsnm{Yamuah}, \binits{L.}},
\bauthor{\bsnm{Delchambre}, \binits{M.}},
\bauthor{\bsnm{Voss}, \binits{G.}},
\bauthor{\bsnm{Greenwood}, \binits{B.M.}},
\bauthor{\bsnm{Hill}, \binits{A.}},
\bauthor{\bsnm{McAdam}, \binits{K.P.}},
\bauthor{\bsnm{Tornieporth}, \binits{N.}},
\bauthor{\bsnm{Cohen}, \binits{J.D.}},
\bauthor{\bsnm{Doherty}, \binits{T.}}:
\batitle{Efficacy of {RTS,S/ASO2} malaria vaccine against {P}lasmodium
  falciparum infection in semi-immune adult men in the {G}ambia: A randomized
  trial}.
\bjtitle{Lancet}
\bvolume{358}(\bissue{9297}),
\bfpage{1927}--\blpage{1934}
(\byear{2001})
\end{barticle}
\endbibitem

\bibitem{asante2011}
\begin{barticle}
\bauthor{\bsnm{Asante}, \binits{K.P.}},
\bauthor{\bsnm{Abdulla}, \binits{S.}},
\bauthor{\bsnm{Agnandji}, \binits{S.}},
\bauthor{\bsnm{Lyimo}, \binits{J.}},
\bauthor{\bsnm{Vekemans}, \binits{J.}},
\bauthor{\bsnm{Soulanoudjingar}, \binits{S.}},
\bauthor{\bsnm{Owuso}, \binits{R.}},
\bauthor{\bsnm{Shomari}, \binits{M.}},
\bauthor{\bsnm{Leach}, \binits{A.}},
\bauthor{\bsnm{Jongert}, \binits{E.}},
\bauthor{\bsnm{Salim}, \binits{N.}},
\bauthor{\bsnm{Frenandes}, \binits{J.F.}},
\bauthor{\bsnm{Dosoo}, \binits{D.}},
\bauthor{\bsnm{Chikawe}, \binits{M.}},
\bauthor{\bsnm{S.}, \binits{I.}},
\bauthor{\bsnm{Osei-Kwakye}, \binits{K.}},
\bauthor{\bsnm{Lievens}, \binits{M.}},
\bauthor{\bsnm{Paricek}, \binits{M.}},
\bauthor{\bsnm{M{\"o}ller}, \binits{T.}},
\bauthor{\bsnm{Apanga}, \binits{S.}},
\bauthor{\bsnm{Mwangoka}, \binits{G.}},
\bauthor{\bsnm{Dubois}, \binits{M.C.}},
\bauthor{\bsnm{Madi}, \binits{T.}},
\bauthor{\bsnm{Kwara}, \binits{E.}},
\bauthor{\bsnm{Minja}, \binits{R.}},
\bauthor{\bsnm{Hounkpatin}, \binits{A.B.}},
\bauthor{\bsnm{Boahen}, \binits{O.}},
\bauthor{\bsnm{Kayan}, \binits{K.}},
\bauthor{\bsnm{Adjedi}, \binits{G.}},
\bauthor{\bsnm{Chandrahoman}, \binits{D.}},
\bauthor{\bsnm{Carter}, \binits{T.}},
\bauthor{\bsnm{Vansadia}, \binits{P.}},
\bauthor{\bsnm{Sillman}, \binits{M.}},
\bauthor{\bsnm{Savarese}, \binits{B.}},
\bauthor{\bsnm{Loucq}, \binits{C.}},
\bauthor{\bsnm{Lapierre}, \binits{D.}},
\bauthor{\bsnm{Greenwood}, \binits{B.}},
\bauthor{\bsnm{Cohen}, \binits{J.}},
\bauthor{\bsnm{Owusu-Ageyi}, \binits{S.}},
\bauthor{\bsnm{Tanner}, \binits{M.}},
\bauthor{\bsnm{Lell}, \binits{B.}}:
\batitle{Safety and efficacy of the {RTS,S/AS01E} candidate malaria vaccine
  given with expanded programme on immunisation vaccines: 19 month follow-up of
  a randomised, open-label, phase 2 trial}.
\bjtitle{Lancet Infectious Diseases}
\bvolume{11}(\bissue{10}),
\bfpage{741}--\blpage{749}
(\byear{2011})
\end{barticle}
\endbibitem

\bibitem{koella+antia2003}
\begin{botherref}
\oauthor{\bsnm{Koella}, \binits{J.C.}},
\oauthor{\bsnm{Antia}, \binits{R.}}:
Epidemiological models for the spread of anti-malarial resistance.
Malaria Journal
\textbf{2}(3)
(2003).
http://www.malariajournal.com/content/2/1/3
\end{botherref}
\endbibitem

\bibitem{dawes+more2009}
\begin{botherref}
\oauthor{\bsnm{Dawes}, \binits{E.J.}},
\oauthor{\bsnm{Churcher}, \binits{T.S.}},
\oauthor{\bsnm{Zhuang}, \binits{A.}},
\oauthor{\bsnm{Sinden}, \binits{R.E.}},
\oauthor{\bsnm{Basanez}, \binits{M.}}:
{Anopheles} mortality is both age and {Plasmodium} density dependent:
  Implications for malaria transmission.
Malaria Journal
\textbf{8}(228)
(2009).
http://www.malariajournal.com/content/8/1/228
\end{botherref}
\endbibitem

\bibitem{koudou+more2014}
\begin{botherref}
\oauthor{\bsnm{Koudou}, \binits{B.G.}},
\oauthor{\bsnm{Malone}, \binits{D.}},
\oauthor{\bsnm{Hemingway}, \binits{J.}}:
The use of motion detectors to estimate net usage by householders, in relation
  to mosquito density in central {C}\^{o}te d'{I}voire: Preliminary results.
Parasites \& Vectors
\textbf{7}(96)
(2014).
doi: 10.1186/1756-3305-7-96
\end{botherref}
\endbibitem

\bibitem{smith+mckenzie2004}
\begin{botherref}
\oauthor{\bsnm{Smith}, \binits{D.L.}},
\oauthor{\bsnm{McKenzie}, \binits{F.E.}}:
Statics and dynamics of malaria infection in {A}nopheles mosquitoes.
Malaria Journal
\textbf{3}(13)
(2004).
http://www.malariajournal.com/content/3/1/13
\end{botherref}
\endbibitem

\bibitem{aron+may1982}
\begin{bchapter}
\bauthor{\bsnm{Aron}, \binits{J.L.}},
\bauthor{\bsnm{May}, \binits{R.M.}}:
\bctitle{The population dynamics of malaria}.
In: \beditor{\bsnm{Anderson}, \binits{R.M.}} (ed.)
\bbtitle{The Population Dynamics of Infectious Diseases: Theory and
  Applications},
pp. \bfpage{168}--\blpage{169}.
\bpublisher{Chapman and Hall},
\blocation{London}
(\byear{1982})
\end{bchapter}
\endbibitem

\bibitem{sutherland+all2005}
\begin{botherref}
\oauthor{\bsnm{Sutherland}, \binits{C.J.}},
\oauthor{\bsnm{Ord}, \binits{R.}},
\oauthor{\bsnm{Dunyo}, \binits{S.}},
\oauthor{\bsnm{Jawara}, \binits{M.}},
\oauthor{\bsnm{Drakeley}, \binits{C.J.}},
\oauthor{\bsnm{Alexander}, \binits{N.}},
\oauthor{\bsnm{Colemand}, \binits{R.}},
\oauthor{\bsnm{Pinder}, \binits{M.}},
\oauthor{\bsnm{Walraven}, \binits{G.}},
\oauthor{\bsnm{Targett}, \binits{G.A.T.}}:
Reduction of malaria transmission to {Anopheles} mosquitoes with a six-dose
  regimen of co-artemether.
{PLoS Medicine}
\textbf{2}(4)
(2005)
\end{botherref}
\endbibitem

\bibitem{white2011}
\begin{botherref}
\oauthor{\bsnm{White}, \binits{M.T.}},
\oauthor{\bsnm{Conteh}, \binits{L.}},
\oauthor{\bsnm{Cibulskis}, \binits{R.}},
\oauthor{\bsnm{Ghani}, \binits{A.C.}}:
Costs and cost-effectiveness of malaria control interventions - a systematic
  review.
Malaria Journal
\textbf{10}(337)
(2011).
http://www.malariajournal.com/content/10/1/337
\end{botherref}
\endbibitem

\bibitem{WHO2014nets}
\begin{botherref}
\oauthor{\bsnm{{World Health Organization}}}:
Estimating population access to {ITNs} versus quantifying for procurement for
  mass campaigns.
Online.
http://www.who.int/malaria/publications/atoz/who-clarification-estimating-population-access-itn-mar2014.pdf
  Accessed {J}anuary 2016.
({M}arch 2014)
\end{botherref}
\endbibitem

\bibitem{seo2014}
\begin{botherref}
\oauthor{\bsnm{Seo}, \binits{M.K.}},
\oauthor{\bsnm{Baker}, \binits{P.}},
\oauthor{\bsnm{Ngoc-Lan}, \binits{K.}}:
Cost-effectiveness analysis of vaccinating children in {M}alawi with {RTS,S}
  vaccines in comparison with long-lasting insecticide-treated nets.
Malaria Journal
\textbf{13}(66)
(2014).
http://www.malariajournal.com/content/13/1/66
\end{botherref}
\endbibitem

\bibitem{BLSCalc}
\begin{botherref}
\oauthor{\bsnm{{United States Bureau of Labor Statistics}}}:
Inflation Calculator.
Online
(Accessed {J}uly 2015).
\url{http://www.bls.gov/data/inflation_calculator.htm}
\end{botherref}
\endbibitem

\bibitem{chitnis+hyman+cushing2008}
\begin{barticle}
\bauthor{\bsnm{Chitnis}, \binits{N.}},
\bauthor{\bsnm{Hyman}, \binits{J.M.}},
\bauthor{\bsnm{Cushing}, \binits{J.M.}}:
\batitle{Determining important parameters in the spread of malaria through the
  sensitivity analysis of a mathematical model}.
\bjtitle{Bulletin of Mathematical Biology}
\bvolume{70}(\bissue{5}),
\bfpage{1272}--\blpage{1296}
(\byear{2008}).
\bcomment{doi: 10.1007/s11538-008-9299-0}
\end{barticle}
\endbibitem

\bibitem{ciafactbook}
\begin{botherref}
\oauthor{\bsnm{{Central Intelligence Agency}}}:
The World Factbook: Kenya.
Online.
https://www.cia.gov/library/publications/the-world-factbook/geos/ke.html
(Accessed {J}une 2014)
\end{botherref}
\endbibitem

\bibitem{filipe+all2007}
\begin{barticle}
\bauthor{\bsnm{Filipe}, \binits{J.A.N.}},
\bauthor{\bsnm{Riley}, \binits{E.M.}},
\bauthor{\bsnm{Drakeley}, \binits{C.J.}},
\bauthor{\bsnm{Sutherland}, \binits{C.J.}},
\bauthor{\bsnm{Ghani}, \binits{A.C.}}:
\batitle{Determination of the processes driving the acquisition of immunity to
  malaria using a mathematical transmission model}.
\bjtitle{PLoS Computational Biology}
\bvolume{3}(\bissue{12}),
\bfpage{2569}--\blpage{2579}
(\byear{2007}).
\bcomment{doi:10.1371/journal.pcbi.0030255}
\end{barticle}
\endbibitem

\bibitem{anderson+may1991}
\begin{bbook}
\bauthor{\bsnm{Anderson}, \binits{R.M.}},
\bauthor{\bsnm{May}, \binits{R.M.}}:
\bbtitle{Infectious Diseases of Humans: Dynamics and Control}.
\bpublisher{Oxford University Press},
\blocation{Oxford}
(\byear{1991}).
\bcomment{pp. 399}
\end{bbook}
\endbibitem

\bibitem{kakkilaya2011}
\begin{botherref}
\oauthor{\bsnm{Kakkilaya}, \binits{B.S.}}:
Malaria Site: Immunity.
(Accessed {July} 2015).
\url{http://www.malariasite.com/immunity/}
\end{botherref}
\endbibitem

\bibitem{zhou+all2010}
\begin{botherref}
\oauthor{\bsnm{Zhou}, \binits{G.}},
\oauthor{\bsnm{Githeko}, \binits{A.K.}},
\oauthor{\bsnm{Minakawa}, \binits{N.}},
\oauthor{\bsnm{Yan}, \binits{G.}}:
Community-wide benefits of targeted indoor residual spray for malaria control
  in the {Western Kenya Highland}.
Malaria Journal
\textbf{9}(67)
(2010).
http://www.malariajournal.com/content/9/1/67
\end{botherref}
\endbibitem

\bibitem{olotu+all2013}
\begin{barticle}
\bauthor{\bsnm{Olotu}, \binits{A.}},
\bauthor{\bsnm{Fegan}, \binits{G.}},
\bauthor{\bsnm{Wambua}, \binits{J.}},
\bauthor{\bsnm{Nyangweso}, \binits{G.}},
\bauthor{\bsnm{Awuondo}, \binits{K.O.}},
\bauthor{\bsnm{Leach}, \binits{A.}},
\bauthor{\bsnm{Lievens}, \binits{M.}},
\bauthor{\bsnm{Leboulleux}, \binits{D.}},
\bauthor{\bsnm{Njuguna}, \binits{P.}},
\bauthor{\bsnm{Peshu}, \binits{N.}},
\bauthor{\bsnm{Marsh}, \binits{K.}},
\bauthor{\bsnm{Bejon}, \binits{P.}}:
\batitle{Four-year efficacy of {RTS,S/AS01E} and its interaction with malaria
  exposure}.
\bjtitle{The {N}ew {E}ngland Journal of Medicine}
\bvolume{368}(\bissue{12}),
\bfpage{1111}--\blpage{1120}
(\byear{2013})
\end{barticle}
\endbibitem

\bibitem{pmikenya2013}
\begin{botherref}
\oauthor{\bsnm{{President's Malaria Initiative}}}:
FY 2013 {K}enya - Revised Funding Table.
Online.
http://www.pmi.gov/where-we-work/kenya
(2013)
\end{botherref}
\endbibitem

\bibitem{WHO-drugresistance}
\begin{botherref}
\oauthor{\bsnm{{World Health Organization}}}:
Antimalarial drug resistance.
Online.
http://www.who.int/malaria/areas/drug\_resistance/overview/en/
\end{botherref}
\endbibitem

\bibitem{kim+schneider2013}
\begin{barticle}
\bauthor{\bsnm{Kim}, \binits{Y.}},
\bauthor{\bsnm{Schneider}, \binits{K.A.}}:
\batitle{Evolution of drug resistance in malaria parasite populations}.
\bjtitle{Nature Education Knowledge}
\bvolume{4}(\bissue{8}),
\bfpage{6}
(\byear{2013}).
\bcomment{http://www.nature.com/scitable/knowledge/library/evolution-of-drug-resistance-in-malaria-parasite-96645809}
\end{barticle}
\endbibitem

\bibitem{edi2012}
\begin{botherref}
\oauthor{\bsnm{Edi}, \binits{V.A.}},
\oauthor{\bsnm{Koudou}, \binits{B.G.}},
\oauthor{\bsnm{Jones}, \binits{C.M.}},
\oauthor{\bsnm{Weetman}, \binits{D.}},
\oauthor{\bsnm{Ranson}, \binits{H.}}:
Multiple-insecticide resistance in {Anopheles gambiae} mosquitoes, southern
  {C}\^{o}te d'{I}voire.
Emerging Infectious Diseases
\textbf{18}(9)
(2012).
http://wwwnc.cdc.gov/eid/article/18/9/12-0262\_article
\end{botherref}
\endbibitem

\bibitem{czyzyk+all1998}
\begin{barticle}
\bauthor{\bsnm{Czyzyk}, \binits{J.}},
\bauthor{\bsnm{Mesnier}, \binits{M.P.}},
\bauthor{\bsnm{Mor\'{e}}, \binits{J.J.}}:
\batitle{The {NEOS} server}.
\bjtitle{IEEE Journal on Computational Science and Engineering}
\bvolume{5}(\bissue{3}),
\bfpage{68}--\blpage{75}
(\byear{1998}).
\bcomment{doi: 10.1109/99.714603}
\end{barticle}
\endbibitem

\bibitem{dolan2001}
\begin{botherref}
\oauthor{\bsnm{Dolan}, \binits{E.}}:
The {NEOS} server 4.0 administrative guide. {T}echnical memorandum
  {ANL/MCS-TM-250}.
Technical report,
Mathematics and Computer Science Division, Argonne National Laboratory
(2001).
http://info.mcs.anl.gov/pub/tech\_reports/reports/TM-250.pdf
\end{botherref}
\endbibitem

\bibitem{gropp+more1997}
\begin{bchapter}
\bauthor{\bsnm{Gropp}, \binits{W.}},
\bauthor{\bsnm{Mor\'{e}}, \binits{J.J.}}:
\bctitle{Optimization environments and the {NEOS} server}.
In: \beditor{\bsnm{Buhmann}, \binits{M.D.}},
\beditor{\bsnm{Iserles}, \binits{A.}} (eds.)
\bbtitle{Approximation Theory and Optimization},
pp. \bfpage{167}--\blpage{182}.
\bpublisher{Cambridge University Press},
\blocation{Cambridge}
(\byear{1997})
\end{bchapter}
\endbibitem

\end{thebibliography}




     

\section*{Tables}

\begin{table}[h!] 
\begin{center}
  \begin{tabular}{| l || c | c | c }
    \hline
    \textbf{Intervention} & \textbf{Cost} \\ \hline \hline
     None & 0  \\ \hline
     LLIN & 1.33  \\ \hline
     ACT & 4.82  \\ \hline
     IPT  & 1.13   \\ \hline
     IRS  & 2.22  \\ \hline
     Vaccine  & 20.66  \\ \hline
  \end{tabular} 
\end{center}
\caption{Baseline cost for using interventions, per treated person for one year, in 2009 USD \cite{white2011, WHO2014nets, seo2014}.}
\label{tab:costs}
\end{table}

\begin{landscape}
\begin{table}[h!] 
\begin{center}
\begin{tabular}{|l || c | c || c | c| c| c| c|}
    \hline
    &  & \textbf{Baseline Value} & \multicolumn{5}{c}{\textbf{Treatment Value}} \\ \cline{4-8}
 \textbf{Symbol}     & \textbf{Description} &  (untreated) & LLIN & IRS& IPT & ACT & Vaccine \\ \hline \hline
    $a_u$, $a_t$ &  Bites per mosquito per human per day & 0.25 \cite{chitnis+hyman+cushing2008} & $a_u(1-\beta)$ \cite{killeen+all2006} & & & & \\ \hline
$b_u$, $b_t$        & Transmission efficacy from infected & 0.022   \cite{chitnis+hyman+cushing2008} & & & 0.0047 & & 0.005  \\
 & mosquito to susceptible, untreated human   & & & & & & \\ \hline
$\beta$ & Proportion of bites that would& & 0.8 \cite{killeen+all2006} & & & &  \\
& occur while sleeping  & &  & & & & \\\hline
$c$     & Transmission efficacy from & 0.36 \cite{chitnis+hyman+cushing2008} & & & & &   \\ 
    & infected human to mosquito& & & & & &  \\ \hline
 $\delta$     & Daily birth rate and death rate   & $4.7895*10^{-5}$ \cite{ciafactbook} & & & & & \\
  & assuming constant population &  & & & & & \\ \hline
    $\gamma_u$, $\gamma_t$ & Recovery rate in humans & $\frac{1}{180}$ \cite{filipe+all2007} & & & & $\frac{1}{10}$ & $\frac{1}{5.5}$\\ \hline
    $m_u$, $m_t$ & Mosquitoes per human & 20 \cite{mandal+sarkar+sinha2011} & $m_u(1-\beta \chi_{LLIN})$ & $m_{u,IRS}=m_u(1-q\chi_{IRS_u})$ & & &  \\
   &  & &  & $m_{t}=m_u(1-\chi_{IRS_t})$ \cite{zhou+all2010}& & &  \\\hline
    $\mu$ & Mosquito mortality rate & 0.095  \cite{anderson+may1991,filipe+all2007} & & & & &  \\ \hline
    $\omega$ & Duration of immunity without reinfection  & 274 days \cite{kakkilaya2011} & & & & & \\ \hline
    $q$ & Treatment coverage & 0.2, 0.4 or 0.6 & & & & &  \\ \hline
    $\tau$ & Incubation period in mosquito & 10 days  \cite{anderson+may1991} & & & & &  \\  \hline
$x$ & Fraction of the population & 0.146  \cite{ciafactbook} & & & & &  \\ 
 &  0-4 years of age  &  & & & & & \\  \hline
$\chi_{LLIN}$  & Probability of mosquito mortality  & & 0.8 \cite{lindblade+all2005} & & & &  \\ 
& when exposed to a treated net &&&&&&  \\ \hline
$\chi_{IRS_t}$  & Percent reduction in mosquito density  & & & 0.95 \cite{zhou+all2010} & & & \\ 
& in a house treated with IRS & & &  & & & \\ \hline
$\chi_{IRS_u}$  & Percent reduction in mosquito density  & & & 0.5 \cite{zhou+all2010}  & & & \\ 
& in an untreated house when all  & & && & & \\ 
 & nearby houses are treated with IRS & & & & & & \\ 
    \hline \hline
\end{tabular}
\end{center}
\caption{Malaria transmission parameter values for the baseline, untreated case (corresponding to the subscript ``$u$'') and treatment cases (corresponding to the subscript ``$t$'').  In the Baseline Value column, an empty space means that the parameter does not apply to the baseline, untreated case.  In the Treatment Value columns, an empty space means that the parameter is unchanged by that particular intervention.}
\label{tab:parameters}
\end{table}
\end{landscape}

\begin{landscape}
\begin{table}[h!]
{\small
\begin{center}
\begin{tabular}{|l | c | c || c | c| c| c| c|}
\hline
Geographic  & Number of  & Initial  & Y1 Intervention & Y2 Intervention & Y3 Intervention & Y4 Intervention & Y5 Intervention \\
Region & Districts & Population State & (End Pop. State) & (End Pop. State) & (End Pop. State)  & (End Pop. State) & (End Pop. State) \\ \hline \hline
(D, L) &	500 &	(60, 15, 25) &	ACT 60\% &	NONE & 	NONE &	NONE &	NONE \\	
&	&	&	(90, 0, 10) &	(95, 0, 5) &	(100, 0, 0) &	(100, 0, 0) &	(100, 0, 0) \\ \hline	
(D, M) &	500 &	(60, 15, 25) &	ACT 60\% &	NONE & 	NONE &	NONE &	NONE \\	
&	&	&	(90, 0, 10) &	(95, 0, 5) &	(100, 0, 0) &	(100, 0, 0) &	(100, 0, 0) \\ \hline	
(D, H) &	500 &	(60, 15, 25) &	ACT 60\% &	NONE & 	NONE &	NONE &	NONE \\	
&	&	&	(90, 0, 10) &	(95, 0, 5) &	(100, 0, 0) &	(100, 0, 0) &	(100, 0, 0) \\ \hline	
(M, L) &	500 &	(15, 15, 70) &	LLIN\_ACT 60\% &	ACT\_IRS 60\% & 	ACT\_IRS 60\% &	ACT\_IRS 60\% &	ACT\_IRS 60\% \\	
&	&	&	(65, 5, 30) &	(80, 5, 15) &	(85, 5, 10) &	(85, 5, 10) &	(85, 5, 10) \\ \hline	
(M, M) &	500 &	(15, 15, 70) &	LLIN\_ACT 60\% &	ACT\_IRS 60\% & 	ACT\_IRS 60\% &	ACT\_IRS 60\% &	LLIN\_ACT 60\% \\	
&	&	&	(65, 5, 30) &	(80, 5, 15) &	(85, 5, 10) &	(85, 5, 10) &	(75, 10, 15) \\ \hline	
(M, H) &	500 &	(15, 15, 70) &	LLIN\_ACT 60\% &	ACT\_IRS 60\% & 	ACT\_IRS 60\% &	ACT\_IRS 60\% &	LLIN\_ACT 60\% \\	
&	&	&	(65, 5, 30) &	(80, 5, 15) &	(85, 5, 10) &	(85, 5, 10) &	(75, 10, 15) \\ \hline	
(W, L) &	395 &	(10, 15, 75) &	LLIN\_ACT 20\% &	ACT\_IRS 60\% & 	LLIN\_ACT 60\% &	LLIN\_ACT 60\% &	LLIN\_ACT 60\% \\	
&	&	&	(25, 10, 65) &	(65, 5, 30) &	(60, 10, 30) &	(60, 10, 30) &	(60, 10, 30) \\ \hline	
(W, L) &	105 &	(10, 15, 75) &	LLIN\_ACT 60\% &	LLIN\_ACT 60\% & 	LLIN\_ACT 60\% &	LLIN\_ACT 60\% &	LLIN\_ACT 60\% \\	
&	&	&	(60, 5, 35) &	(60, 10, 30) &	(60, 10, 30) &	(60, 10, 30) &	(60, 10, 30) \\ \hline	
(W, M) &	276 &	(10, 15, 75) &	NONE &	LLIN\_ACT 60\% & 	LLIN\_ACT 60\% &	LLIN\_ACT 60\% &	LLIN\_ACT 60\% \\	
&	&	&	(10, 15, 75) &	(60, 5, 35) &	(60, 10, 30) &	(60, 10, 30) &	(60, 10, 30) \\ \hline	
(W, M) &	136 &	(10, 15, 75) &	NONE &	LLIN\_ACT 20\% & 	ACT\_IRS 60\% &	LLIN\_ACT 60\% &	LLIN\_ACT 60\% \\	
&	&	&	(10, 15, 75) &	(25, 10, 65) &	(65, 5, 30) &	(60, 10, 30) &	(60, 10, 30) \\ \hline	
(W, M) &	59 &	(10, 15, 75) &	NONE &	NONE & 	LLIN\_ACT 60\% &	LLIN\_ACT 60\% &	LLIN\_ACT 60\% \\	
&	&	&	(10, 15, 75) &	(10, 15, 75) &	(60, 5, 35) &	(60, 10, 30) &	(60, 10, 30) \\ \hline	
(W, M) &	22 &	(10, 15, 75) &	NONE &	NONE & 	NONE &	LLIN 20\% &	LLIN\_ACT 60\% \\	
&	&	&	(10, 15, 75) &	(10, 15, 75) &	(10, 15, 75) &	(25, 10, 65) &	(55, 10, 35) \\ \hline	
(W, M) &	3 &	(10, 15, 75) &	NONE &	NONE & 	IPT 40\% &	LLIN 20\% &	LLIN\_ACT 60\% \\	
&	&	&	(10, 15, 75) &	(10, 15, 75) &	(10, 15, 75) &	(25, 10, 65) &	(55, 10, 35) \\ \hline	
(W, M) &	2 &	(10, 15, 75) &	NONE &	IPT 40\% & 	LLIN\_ACT 60\% &	LLIN\_ACT 60\% &	LLIN\_ACT 60\% \\	
&	&	&	(10, 15, 75) &	(10, 15, 75) &	(60, 5, 35) &	(60, 10, 30) &	(60, 10, 30) \\ \hline	
(W, M) &	1 &	(10, 15, 75) &	IPT 20\% &	LLIN\_ACT 60\% & 	LLIN\_ACT 60\% &	LLIN\_ACT 60\% &	LLIN\_ACT 60\% \\	
&	&	&	(10, 15, 75) &	(60, 5, 35) &	(60, 10, 30) &	(60, 10, 30) &	(60, 10, 30) \\ \hline	
(W, M) &	1 &	(10, 15, 75) &	IPT 40\% &	LLIN\_ACT 60\% & 	LLIN\_ACT 60\% &	LLIN\_ACT 60\% &	LLIN\_ACT 60\% \\	
&	&	&	(10, 15, 75) &	(60, 5, 35) &	(60, 10, 30) &	(60, 10, 30) &	(60, 10, 30) \\ \hline	
(W, H) &	371 &	(10, 15, 75) &	NONE &	NONE & 	NONE &	NONE &	LLIN\_ACT 60\% \\	
&	&	&	(10, 15, 75) &	(10, 15, 75) &	(10, 15, 75) &	(10, 15, 75) &	(60, 5, 35) \\ \hline	
(W, H) &	126 &	(10, 15, 75) &	NONE &	NONE & 	NONE &	NONE &	NONE \\	
&	&	&	(10, 15, 75) &	(10, 15, 75) &	(10, 15, 75) &	(10, 15, 75) &	(10, 15, 75) \\ \hline	
(W, H) &	1 &	(10, 15, 75) &	NONE &	NONE & 	NONE &	NONE &	IPT 40\% \\	
&	&	&	(10, 15, 75) &	(10, 15, 75) &	(10, 15, 75) &	(10, 15, 75) &	(10, 15, 75) \\ \hline	
(W, H) &	1 &	(10, 15, 75) &	NONE &	NONE & 	IPT 40\% &	NONE &	LLIN\_ACT 60\% \\	
&	&	&	(10, 15, 75) &	(10, 15, 75) &	(10, 15, 75) &	(10, 15, 75) &	(60, 5, 35) \\ \hline	
(W, H) &	1 &	(10, 15, 75) &	IPT 40\% &	IPT 40\% & 	IPT 40\% &	IPT 20\% &	LLIN\_ACT 60\% \\	
&	&	&	(10, 15, 75) &	(10, 15, 75) &	(10, 15, 75) &	(10, 15, 75) &	(60, 5, 35) \\ \hline	
\hline
\end{tabular}
\end{center}
\caption{Optimal five-year (Y1 through Y5) sequences of interventions for the baseline efficacy scenario.  Geographic Region refers to the climate region and distribution cost pair, where the climate region is dry (D), moderate (M) or wet (W), and the distribution cost is low (L), medium (M) or high (H).  Number of Districts refers to the number of districts that were assigned a given trajectory.  Initial Population State is the starting $(S,I,R)$ percentages of the region; the End Population State for a given year is the resulting $(S,I,R)$ state after distributing the corresponding intervention.  The total person-days of malaria infection in this scenario is 4.506 billion. 
	}
\label{tab:baselineoutput}}
\end{table}
\end{landscape}

\begin{landscape}
\begin{table}[h!]
\begin{center}
\begin{tabular}{|l | c | c || c | c| c| c| c|}
\hline
Geographic  & Number of  & Initial  & Y1 Intervention & Y2 Intervention & Y3 Intervention & Y4 Intervention & Y5 Intervention \\
Region & Districts & Population State & (End Pop. State) & (End Pop. State) & (End Pop. State)  & (End Pop. State) & (End Pop. State) \\ \hline \hline
(D, L) &	500 &	(60, 15, 25) &	LLIN\_ACT  60\% &	ACT 60\% & 	ACT 20\% &	ACT 60\% &	NONE \\	
&	&	&	(90, 0, 10) &	(95, 0, 5) &	(100, 0, 0) &	(100, 0, 0) &	(100, 0, 0) \\ \hline	
(D, M) &	500 &	(60, 15, 25) &	LLIN\_ACT  60\% &	ACT 40\% & 	ACT 40\% &	ACT 20\% &	NONE \\	
&	&	&	(90, 0, 10) &	(95, 0, 5) &	(100, 0, 0) &	(100, 0, 0) &	(100, 0, 0) \\ \hline	
(D, H) &	500 &	(60, 15, 25) &	ACT 60\% &	ACT 60\% & 	ACT 40\% &	ACT 60\% &	ACT 40\% \\	
&	&	&	(85, 5, 10) &	(95, 0, 5) &	(100, 0, 0) &	(100, 0, 0) &	(100, 0, 0) \\ \hline	
(M, L) &	500 &	(15, 15, 70) &	LLIN\_ACT  60\% &	ACT\_IRS 60\% & 	NONE &	ACT 40\% &	NONE \\	
&	&	&	(65, 5, 30) &	(90, 0, 10) &	(95, 0, 5) &	(100, 0, 0) &	(100, 0, 0) \\ \hline	
(M, M) &	500 &	(15, 15, 70) &	LLIN\_ACT  60\% &	ACT\_IRS 60\% & 	NONE &	ACT 20\% &	ACT 40\% \\	
&	&	&	(65, 5, 30) &	(90, 0, 10) &	(95, 0, 5) &	(100, 0, 0) &	(100, 0, 0) \\ \hline	
(M, H) &	266 &	(15, 15, 70) &	LLIN\_ACT  60\% &	ACT\_IRS 60\% & 	ACT 60\% &	NONE &	ACT 40\% \\	
&	&	&	(65, 5, 30) &	(90, 0, 10) &	(95, 0, 5) &	(100, 0, 0) &	(100, 0, 0) \\ \hline	
(M, H) &	234 &	(15, 15, 70) &	LLIN 60\% &	ACT\_IRS 60\% & 	ACT 60\% &	NONE &	ACT 40\% \\	
&	&	&	(60, 5, 35) &	(90, 0, 10) &	(95, 0, 5) &	(100, 0, 0) &	(100, 0, 0) \\ \hline	
(W, L) &	497 &	(10, 15, 75) &	NONE &	LLIN\_ACT  60\% & 	ACT\_IRS 60\% &	ACT\_IRS 60\% &	ACT\_IRS 60\% \\	
&	&	&	(10, 15, 75) &	(60, 5, 35) &	(80, 5, 15) &	(75, 10, 15) &	(75, 10, 15) \\ \hline	
(W, L) &	3 &	(10, 15, 75) &	IPT 40\% &	LLIN\_ACT  60\% & 	ACT\_IRS 60\% &	ACT\_IRS 60\% &	ACT\_IRS 60\% \\	
&	&	&	(10, 15, 75) &	(60, 5, 35) &	(80, 5, 15) &	(75, 10, 15) &	(75, 10, 15) \\ \hline	
(W, M) &	324 &	(10, 15, 75) &	NONE &	LLIN\_ACT  60\% & 	ACT\_IRS 60\% &	ACT\_IRS 60\% &	ACT\_IRS 60\% \\	
&	&	&	(10, 15, 75) &	(60, 5, 35) &	(80, 5, 15) &	(75, 10, 15) &	(75, 10, 15) \\ \hline	
(W, M) &	172 &	(10, 15, 75) &	NONE &	NONE & 	LLIN\_ACT  60\% &	ACT\_IRS 60\% &	ACT\_IRS 60\% \\	
&	&	&	(10, 15, 75) &	(10, 15, 75) &	(60, 5, 35) &	(80, 5, 15) &	(75, 10, 15) \\ \hline	
(W, M) &	2 &	(10, 15, 75) &	IPT 40\% &	LLIN\_ACT  60\% & 	ACT\_IRS 60\% &	ACT\_IRS 60\% &	ACT\_IRS 60\% \\	
&	&	&	(10, 15, 75) &	(60, 5, 35) &	(80, 5, 15) &	(75, 10, 15) &	(75, 10, 15) \\ \hline	
(W, M) &	1 &	(10, 15, 75) &	NONE &	IPT 20\% & 	LLIN\_ACT  60\% &	ACT\_IRS 60\% &	ACT\_IRS 60\% \\	
&	&	&	(10, 15, 75) &	(10, 15, 75) &	(60, 5, 35) &	(80, 5, 15) &	(75, 10, 15) \\ \hline	
(W, M) &	1 &	(10, 15, 75) &	NONE &	IPT 40\% & 	LLIN\_ACT  60\% &	ACT\_IRS 60\% &	ACT\_IRS 60\% \\	
&	&	&	(10, 15, 75) &	(10, 15, 75) &	(60, 5, 35) &	(80, 5, 15) &	(75, 10, 15) \\ \hline	
(W, H) &	498 &	(10, 15, 75) &	NONE &	NONE & 	LLIN\_ACT  60\% &	ACT\_IRS 60\% &	ACT\_IRS 60\% \\	
&	&	&	(10, 15, 75) &	(10, 15, 75) &	(60, 5, 35) &	(80, 5, 15) &	(75, 10, 15) \\ \hline	
(W, H) &	1 &	(10, 15, 75) &	IPT 40\% &	NONE & 	LLIN\_ACT  60\% &	ACT\_IRS 60\% &	ACT\_IRS 60\% \\	
&	&	&	(10, 15, 75) &	(10, 15, 75) &	(60, 5, 35) &	(80, 5, 15) &	(75, 10, 15) \\ \hline	
(W, H) &	1 &	(10, 15, 75) &	IPT 40\% &	IPT 40\% & 	LLIN\_ACT  60\% &	ACT\_IRS 60\% &	ACT\_IRS 60\% \\	
&	&	&	(10, 15, 75) &	(10, 15, 75) &	(60, 5, 35) &	(80, 5, 15) &	(75, 10, 15) \\ \hline	
\hline
\end{tabular}
\end{center}
\caption{Optimal five-year (Y1 through Y5) sequences of interventions for the optimistic efficacy scenario.  Geographic Region refers to the climate region and distribution cost pair, where the climate region is dry (D), moderate (M) or wet (W), and the distribution cost is low (L), medium (M) or high (H).  Number of Districts refers to the number of districts that were assigned a given trajectory.  Initial Population State is the starting $(S,I,R)$ percentages of the region; the End Population State for a given year is the resulting $(S,I,R)$ state after distributing the corresponding intervention.  The total person-days of malaria infection in this scenario is 2.977 billion.
}
\label{tab:optimisticoutput}
\end{table}
\end{landscape}

\begin{landscape}
\begin{table}[h!]
{\small
\begin{center}
\begin{tabular}{|l | c | c || c | c| c| c| c|}
\hline
Geographic  & Number of  & Initial  & Y1 Intervention & Y2 Intervention & Y3 Intervention & Y4 Intervention & Y5 Intervention \\
Region & Districts & Population State & (End Pop. State) & (End Pop. State) & (End Pop. State)  & (End Pop. State) & (End Pop. State) \\ \hline \hline
(D, L) &	500 &	(60, 15, 25) &	ACT 60\% &	NONE & 	NONE &	NONE &	NONE \\	
&	&	&	(90, 0, 10) &	(95, 0, 5) &	(100, 0, 0) &	(100, 0, 0) &	(100, 0, 0) \\ \hline	
(D, M) &	500 &	(60, 15, 25) &	ACT 60\% &	NONE & 	NONE &	NONE &	NONE \\	
&	&	&	(90, 0, 10) &	(95, 0, 5) &	(100, 0, 0) &	(100, 0, 0) &	(100, 0, 0) \\ \hline	
(D, H) &	500 &	(60, 15, 25) &	ACT 60\% &	NONE & 	NONE &	NONE &	NONE \\	
&	&	&	(90, 0, 10) &	(95, 0, 5) &	(100, 0, 0) &	(100, 0, 0) &	(100, 0, 0) \\ \hline	
(M, L) &	494 &	(15, 15, 70) &	ACT 60\% &	ACT\_IRS 60\% & 	ACT\_IRS 60\% &	LLIN\_ACT 60\% &	LLIN\_ACT 60\% \\	
&	&	&	(60, 5, 35) &	(80, 5, 15) &	(75, 10, 15) &	(70, 10, 20) &	(70, 10, 20) \\ \hline	
(M, L) &	6 &	(15, 15, 70) &	ACT 60\% &	ACT\_IRS 60\% & 	LLIN\_ACT 60\% &	LLIN\_ACT 60\% &	LLIN\_ACT 60\% \\	
&	&	&	(60, 5, 35) &	(80, 5, 15) &	(75, 10, 15) &	(70, 10, 20) &	(70, 10, 20) \\ \hline	
(M, M) &	500 &	(15, 15, 70) &	LLIN\_ACT 60\% &	LLIN\_ACT 60\% & 	LLIN\_ACT 60\% &	LLIN\_ACT 60\% &	LLIN\_ACT 60\% \\	
&	&	&	(65, 5, 30) &	(75, 10, 15) &	(70, 10, 20) &	(70, 10, 20) &	(70, 10, 20) \\ \hline	
(M, H) &	500 &	(15, 15, 70) &	LLIN\_ACT 60\% &	LLIN\_ACT 60\% & 	LLIN\_ACT 60\% &	LLIN\_ACT 60\% &	LLIN\_ACT 60\% \\	
&	&	&	(65, 5, 30) &	(75, 10, 15) &	(70, 10, 20) &	(70, 10, 20) &	(70, 10, 20) \\ \hline	
(W, L) &	464 &	(10, 15, 75) &	NONE &	LLIN\_ACT 60\% & 	LLIN\_ACT 60\% &	LLIN\_ACT 60\% &	LLIN\_ACT 60\% \\	
&	&	&	(10, 15, 75) &	(60, 5, 35) &	(60, 10, 30) &	(60, 10, 30) &	(60, 10, 30) \\ \hline	
(W, L) &	35 &	(10, 15, 75) &	LLIN\_ACT 60\% &	LLIN\_ACT 60\% & 	LLIN\_ACT 60\% &	LLIN\_ACT 60\% &	LLIN\_ACT 60\% \\	
&	&	&	(60, 5, 35) &	(60, 10, 30) &	(60, 10, 30) &	(60, 10, 30) &	(60, 10, 30) \\ \hline	
(W, L) &	1 &	(10, 15, 75) &	IPT 60\% &	LLIN\_ACT 60\% & 	LLIN\_ACT 60\% &	LLIN\_ACT 60\% &	LLIN\_ACT 60\% \\	
&	&	&	(10, 15, 75) &	(60, 5, 35) &	(60, 10, 30) &	(60, 10, 30) &	(60, 10, 30) \\ \hline	
(W, M) &	499 &	(10, 15, 75) &	NONE &	LLIN\_ACT 60\% & 	LLIN\_ACT 60\% &	LLIN\_ACT 60\% &	LLIN\_ACT 60\% \\	
&	&	&	(10, 15, 75) &	(60, 5, 35) &	(60, 10, 30) &	(60, 10, 30) &	(60, 10, 30) \\ \hline	
(W, M) &	1 &	(10, 15, 75) &	IPT 40\% &	LLIN\_ACT 60\% & 	LLIN\_ACT 60\% &	LLIN\_ACT 60\% &	LLIN\_ACT 60\% \\	
&	&	&	(10, 15, 75) &	(60, 5, 35) &	(60, 10, 30) &	(60, 10, 30) &	(60, 10, 30) \\ \hline	
(W, H) &	241 &	(10, 15, 75) &	NONE &	LLIN\_ACT 60\% & 	LLIN\_ACT 60\% &	LLIN\_ACT 60\% &	LLIN\_ACT 60\% \\	
&	&	&	(10, 15, 75) &	(60, 5, 35) &	(60, 10, 30) &	(60, 10, 30) &	(60, 10, 30) \\ \hline	
(W, H) &	140 &	(10, 15, 75) &	NONE &	NONE & 	LLIN 20\% &	NONE &	NONE \\	
&	&	&	(10, 15, 75) &	(10, 15, 75) &	(20, 10, 70) &	(5, 15, 80) &	(10, 15, 75) \\ \hline	
(W, H) &	109 &	(10, 15, 75) &	NONE &	LLIN 20\% & 	NONE &	NONE &	NONE \\	
&	&	&	(10, 15, 75) &	(20, 10, 70) &	(5, 15, 80) &	(10, 15, 75) &	(10, 15, 75) \\ \hline	
(W, H) &	3 &	(10, 15, 75) &	LLIN 20\% &	NONE & 	IPT 40\% &	LLIN 20\% &	LLIN\_ACT 60\% \\	
&	&	&	(20, 10, 70) &	(5, 15, 80) &	(10, 15, 75) &	(20, 10, 70) &	(55, 10, 35) \\ \hline	
(W, H) &	2 &	(10, 15, 75) &	NONE &	IPT 60\% & 	LLIN 20\% &	NONE &	NONE \\	
&	&	&	(10, 15, 75) &	(10, 15, 75) &	(20, 10, 70) &	(5, 15, 80) &	(10, 15, 75) \\ \hline	
(W, H) &	2 &	(10, 15, 75) &	NONE &	LLIN 20\% & 	NONE &	IPT 40\% &	IPT 60\% \\	
&	&	&	(10, 15, 75) &	(20, 10, 70) &	(5, 15, 80) &	(10, 15, 75) &	(10, 15, 75) \\ \hline	
(W, H) &	1 &	(10, 15, 75) &	NONE &	NONE & 	IPT 60\% &	LLIN 20\% &	IPT 20\% \\	
&	&	&	(10, 15, 75) &	(10, 15, 75) &	(10, 15, 75) &	(20, 10, 70) &	(10, 15, 75) \\ \hline	
(W, H) &	1 &	(10, 15, 75) &	NONE &	IPT 60\% & 	LLIN 20\% &	NONE &	LLIN\_ACT 20\% \\	
&	&	&	(10, 15, 75) &	(10, 15, 75) &	(20, 10, 70) &	(5, 15, 80) &	(25, 10, 65) \\ \hline	
(W, H) &	1 &	(10, 15, 75) &	NONE &	LLIN 20\% & 	NONE &	IPT 40\% &	NONE \\	
&	&	&	(10, 15, 75) &	(20, 10, 70) &	(5, 15, 80) &	(10, 15, 75) &	(10, 15, 75) \\ \hline	

\hline
\end{tabular}
\end{center}
\caption{Optimal five-year (Y1 through Y5) sequences of interventions for the pessimistic efficacy scenario.  Geographic Region refers to the climate region and distribution cost pair, where the climate region is dry (D), moderate (M) or wet (W), and the distribution cost is low (L), medium (M) or high (H).  Number of Districts refers to the number of districts that were assigned a given trajectory.  Initial Population State is the starting $(S,I,R)$ percentages of the region; the End Population State for a given year is the resulting $(S,I,R)$ state after distributing the corresponding intervention.   The total person-days of malaria infection in this scenario is 5.080 billion. 
}
\label{tab:pessimisticoutput}}
\end{table}
\end{landscape}

\begin{landscape}
\begin{table}[h!]
\begin{center}
\begin{tabular}{|l | c | c || c | c| c| c| c|}
\hline
Geographic  & Number of  & Initial  & Y1 Intervention & Y2 Intervention & Y3 Intervention & Y4 Intervention & Y5 Intervention \\
Region & Districts & Population State & (End Pop. State) & (End Pop. State) & (End Pop. State)  & (End Pop. State) & (End Pop. State) \\ \hline \hline
(D, L) &	500 &	(60, 15, 25) &	 ACT 80\% &	 ACT 60\% & 	 NONE &	 ACT 40\% &	 ACT 40\% \\
&	&	&	(90, 0, 10) &	(95, 0, 5) &	(100, 0, 0) &	(100, 0, 0) &	(100, 0, 0) \\ \hline
(D, M) &	500 &	(60, 15, 25) &	 ACT 80\% &	 ACT 80\% & 	 ACT 80\% &	 ACT 40\% &	 ACT 60\% \\
&	&	&	(90, 0, 10) &	(95, 0, 5) &	(100, 0, 0) &	(100, 0, 0) &	(100, 0, 0) \\ \hline
(D, H) &	500 &	(60, 15, 25) &	 ACT 80\% &	 NONE & 	 ACT 80\% &	 ACT 40\% &	 NONE \\
&	&	&	(90, 0, 10) &	(95, 0, 5) &	(100, 0, 0) &	(100, 0, 0) &	(100, 0, 0) \\ \hline
(M, L) &	500 &	(15, 15, 70) &	 ACT 80\% &	 NONE & 	 ACT 80\% &	 ACT 80\% &	 NONE \\
&	&	&	(75, 0, 25) &	(95, 0, 5) &	(100, 0, 0) &	(100, 0, 0) &	(100, 0, 0) \\ \hline
(M, M) &	500 &	(15, 15, 70) &	 ACT 80\% &	 ACT 40\% & 	 ACT 60\% &	 ACT 80\% &	 ACT 40\% \\
&	&	&	(75, 0, 25) &	(95, 0, 5) &	(100, 0, 0) &	(100, 0, 0) &	(100, 0, 0) \\ \hline
(M, H) &	404 &	(15, 15, 70) &	 IPT 40\% &	 LLIN\_ACT 80\% & 	 ACT 40\% &	 ACT 80\% &	 NONE \\
&	&	&	(20, 15, 65) &	(80, 0, 20) &	(95, 0, 5) &	(100, 0, 0) &	(100, 0, 0) \\ \hline
(M, H) &	92 &	(15, 15, 70) &	 ACT 80\% &	 ACT 60\% & 	 ACT 40\% &	 NONE &	 NONE \\
&	&	&	(75, 0, 25) &	(95, 0, 5) &	(100, 0, 0) &	(100, 0, 0) &	(100, 0, 0) \\ \hline
(M, H) &	3 &	(15, 15, 70) &	 NONE &	 LLIN\_ACT 80\% & 	 ACT 40\% &	 ACT 80\% &	 NONE \\
&	&	&	(15, 20, 65) &	(80, 0, 20) &	(95, 0, 5) &	(100, 0, 0) &	(100, 0, 0) \\ \hline
(W, L) &	500 &	(10, 15, 75) &	 LLIN\_ACT 80\% &	 NONE & 	 ACT 40\% &	 ACT 80\% &	 ACT 80\% \\
&	&	&	(75, 0, 25) &	(95, 0, 5) &	(100, 0, 0) &	(100, 0, 0) &	(100, 0, 0) \\ \hline
(W, M) &	500 &	(10, 15, 75) &	 NONE &	 ACT\_IRS 80\% & 	 NONE &	 ACT 60\% &	 ACT 60\% \\
&	&	&	(10, 15, 75) &	(80, 0, 20) &	(95, 0, 5) &	(100, 0, 0) &	(100, 0, 0) \\ \hline
(W, H) &	354 &	(10, 15, 75) &	 NONE &	 ACT\_IRS 80\% & 	 ACT 40\% &	 ACT 40\% &	 NONE \\
&	&	&	(10, 15, 75) &	(80, 0, 20) &	(95, 0, 5) &	(100, 0, 0) &	(100, 0, 0) \\ \hline
(W, H) &	146 &	(10, 15, 75) &	 NONE &	 LLIN\_ACT 80\% & 	 NONE &	 ACT 40\% &	 NONE \\
&	&	&	(10, 15, 75) &	(75, 0, 25) &	(95, 0, 5) &	(100, 0, 0) &	(100, 0, 0) \\ \hline
\hline
\end{tabular}
\end{center}
\caption{Optimal five-year (Y1 through Y5) sequences of interventions for the baseline efficacy scenario when the maximum coverage available for each intervention is 80\%.  Geographic Region refers to the climate region and distribution cost pair, where the climate region is dry (D), moderate (M) or wet (W), and the distribution cost is low (L), medium (M) or high (H).  Number of Districts refers to the number of districts that were assigned a given trajectory.  Initial Population State is the starting $(S,I,R)$ percentages of the region; the End Population State for a given year is the resulting $(S,I,R)$ state after distributing the corresponding intervention.  The total person-days of malaria infection in this scenario is 1.139 billion, and we see that in all geographic regions, malaria is eradicated over the five-year time horizon. 
}
\label{tab:baselinehighcoverage}
\end{table}
\end{landscape}




\end{document}